\documentclass[11pt,letterpaper]{article}

\usepackage[margin=1in]{geometry}

\usepackage{amsmath,amsfonts,amssymb,amsthm}

\usepackage{graphicx}
\usepackage{subcaption}
\usepackage{float}
\usepackage{multirow}
\usepackage{algorithm}
\usepackage{algpseudocode}

\usepackage{tikz}
\usepackage{gensymb}
\usepackage{xspace,bm}
\usepackage{booktabs}
\usepackage{hyperref}
\usepackage{authblk}
\usepackage{cite}

\newcommand{\xmath}[1] {\ensuremath{#1}\xspace}
\newcommand{\blmath}[1] {\xmath{\bm{#1}}}
\newcommand{\x} {\blmath{x}} 
\newcommand{\y} {\blmath{y}}
\newcommand{\m} {\blmath{M}}
\newcommand{\T} {\ensuremath{\bm{\theta}}\xspace}

\title{
Scan-Adaptive Dynamic MRI Undersampling
Using a Dictionary of Efficiently Learned Patterns
}

\author[1]{Siddhant Gautam}
\author[1]{Angqi Li}
\author[2]{Prachi P. Agarwal}
\author[2]{Anil K. Attili}
\author[2,3,4]{Jeffrey A. Fessler}
\author[2,4]{Nicole Seiberlich}
\author[1,5]{Saiprasad Ravishankar}

\affil[1]{Department of Computational Mathematics, Science and Engineering,
Michigan State University, East Lansing, MI, USA}

\affil[2]{Department of Radiology,
University of Michigan, Ann Arbor, MI, USA}

\affil[3]{Department of Electrical Engineering and Computer Science,
University of Michigan, Ann Arbor, MI, USA}

\affil[4]{Department of Biomedical Engineering,
University of Michigan, Ann Arbor, MI, USA}

\affil[5]{Department of Biomedical Engineering,
Michigan State University, East Lansing, MI, USA}

\date{}  

\begin{document}

\maketitle


\begin{abstract}
Cardiac MRI is limited by long acquisition times,
which can lead to patient discomfort and motion artifacts.
We aim to accelerate Cartesian dynamic cardiac MRI
by learning efficient, scan-adaptive undersampling patterns that preserve diagnostic image quality. We develop a learning-based framework for designing scan- or slice-adaptive Cartesian undersampling masks tailored to dynamic cardiac MRI.
Undersampling patterns are optimized using fully sampled training dynamic time-series data. At inference time, a nearest-neighbor search in low-frequency $k$-space selects an optimized mask from a dictionary of learned patterns. Our learned sampling approach improves reconstruction quality across multiple acceleration factors on public and in-house cardiac MRI datasets, including PSNR gains of 2-3 dB, reduced NMSE, improved SSIM, and higher radiologist ratings. The proposed scan-adaptive sampling framework enables faster and higher-quality dynamic cardiac MRI by adapting $k$-space sampling to individual scans.
\end{abstract}



\section{Introduction}\label{sec:introduction}

Magnetic resonance imaging (MRI) provides high-resolution anatomical and functional information with excellent soft-tissue contrast and no ionizing radiation, making it particularly valuable for cardiovascular imaging. However, MRI acquisition is slow, which is especially limiting for dynamic applications such as cardiac cine MRI, where long scan times increase motion sensitivity and reduce clinical throughput. Thus, it is important to design efficient undersampling techniques that balance both acquisition speed and image reconstruction quality. This design depends on selecting a subset of $k$-space samples through an effective undersampling pattern and a dynamic reconstruction algorithm capable of recovering high-quality, temporally consistent images from incomplete data. 

Early MRI acceleration relied on pulse sequence design~\cite{liang2000principles, bernstein2004handbook, tsao2010ultrafast} and parallel imaging (p-MRI)~\cite{pruessmann2006encoding, ying2010parallel,deshmane2012parallel}, which uses multi-coil sensitivity encoding but can suffer from noise amplification at high acceleration. Compressed sensing (CS) addresses this regime by reconstructing from undersampled measurements using sparsity-based regularization~\cite{donoho2006compressed,emmanuel2006robust, lustig2007sparse, lustig2008compressed}.
The earliest work for dynamic MRI reconstruction used the temporal redundancy across frames, motivating methods based on temporal transform sparsity, such as $k$-$t$ BLAST and $k$-$t$ SENSE~\cite{tsao2003k, jung2007improved}, $k$-$t$ ISD~\cite{liang2012k}, and $k$-$t$ SPARSE-SENSE~\cite{feng2013highly}.
Later works incorporated a low rank prior into the reconstruction framework that included $k$-$t$ SLR~\cite{lingala2011accelerated} and L+S~\cite{otazo2015low}.
More recent dynamic MRI methods exploit manifold/low-rank structure in $k$-space,
including methods like SToRM~\cite{poddar2015dynamic}, KLR~\cite{nakarmi2017kernel}, bilinear manifold~\cite{shetty2019bi}, and LTSA~\cite{djebra2022manifold}.

Unrolling-based deep learning techniques have also been used for dynamic reconstruction, some of which include deep CNN cascade~\cite{schlemper2017deep}, convolutional recurrent neural networks (CRNN) for dynamic reconstruction~\cite{qin2018convolutional}, and DIMENSION~\cite{wang2022dimension}. Later, SLR-Net~\cite{ke2021learned} used a learned low rank prior to explore the temporal redundancy in dynamic MRI reconstruction using a learned singular value thresholding operator.
Learned DC~\cite{cheng2021learning} used a deep learning-based approach that implicitly learns the data consistency with deep networks for dynamic MR imaging. 
CTFNet~\cite{qin2021complementary} proposed a recurrent neural network–based framework that exploits complementary regularization in the spatiotemporal (x–t) and temporal Fourier (x–f) domains.

Recent advances have explored implicit neural networks that map spatial and temporal coordinates to image intensities for cardiac MRI reconstruction. Earlier works such as NeRP~\cite{shen2022nerp}, neural implicit k-space~\cite{huang2023neural,huang2024subspace}, and Fourier-feature-based models~\cite{kunz2024implicit} show how implicit representations can reconstruct dynamic cardiac motion directly from sparse measurements. Subsequent methods introduced subspace constraints~\cite{huang2024subspace}, spatiotemporal coordinate encoding~\cite{feng2025spatiotemporal}, and self-supervised or physics-informed training~\cite{spieker2025pisco, catalan2025unsupervised} to improve reconstruction quality and generalization further.

In accelerated MRI, acquisition design plays a critical role in determining image quality. 
Standard compressed sensing (CS) techniques typically rely on fixed heuristic patterns such as variable-density random sampling (VDRS)~\cite{lustig2007sparse}, equispaced sampling~\cite{haldar2010compressed}, Poisson-disc sampling~\cite{bridson2007fast}, and hybrid variable-density Poisson-disc masks~\cite{lustig2010spirit, levine20173d} to promote incoherent aliasing. 
Moving beyond these fixed designs, early optimization techniques derived sampling patterns from static dataset statistics, utilizing metrics such as power spectra and experimental design criteria~\cite{knoll2011adapted, ravishankar2011adaptive, haldar2019oedipus, seeger2010optimization}.

Later active learning frameworks, such as greedy algorithms~\cite{gozcu2018learning,gozcu2019rethinking, sanchez2020scalable} were used to minimize reconstruction error directly. Recent work has focused on algorithmic efficiency, utilizing stochastic relaxation and scalable subset selection methods like BASS~\cite{zibetti2021fast, zibetti2022alternating} to generate population-adaptive masks for large-scale training in multi-coil settings. More recent efforts have explored differentiable pipelines, such as LOUPE~\cite{bahadir2020deep, zhang2020extending} and J-MoDL~\cite{aggarwal2020j}, which jointly optimize sampling and reconstruction end-to-end.

While these methods have primarily been applied to static MRI, dynamic cardiac cine MRI presents unique sampling challenges, requiring patterns that preserve spatiotemporal consistency and cardiac periodicity. Current solutions, such as Variable-density Incoherent Spatiotemporal Acquisition (VISTA)~\cite{ahmad2015variable} rely on fixed heuristics, which utilize fixed trajectories to promote temporal incoherence across all subjects.
In contrast, learnable population-adaptive frameworks like LOUPE~\cite{bahadir2020deep} and J-MoDL~\cite {aggarwal2020j} optimize masks directly from data, but these have been primarily restricted to static imaging. 
Both these approaches share a fundamental limitation as they apply a fixed or population adaptive pattern that ignores the distinct cardiac dynamics and anatomical variations of the individual patient.

On the other hand, scan-adaptive techniques such as SeqMRI~\cite{yin2021end}, MNet~\cite{huang2022single}, SUNO~\cite{gautam2024patient, dhar2025learning, gautam2026scan} tailor acquisition to the individual subject. For instance, SeqMRI formulates sampling as a sequential decision process, optimizing a policy to select $k$-space lines iteratively based on the current reconstruction state. Similarly, MNet uses a CNN to map low-frequency $k$-space data directly to a scan-specific Cartesian mask. While effective for static imaging, these methods face significant barriers in dynamic cardiac MRI. The high dimensionality of dynamic $k$-space renders sequential decision-making and direct mask prediction computationally prohibitive or unstable for ensuring spatiotemporal consistency across cardiac phases.

To address this, we propose Dynamic Scan-Adaptive MRI Undersampling using Neighbor-based Optimization (dSUNO), which extends our prior SUNO framework~\cite{gautam2026scan} to dynamic cardiac MRI. The proposed dSUNO framework replaces the original ICD-based mask optimization with a novel Randomized Batched Iterative Coordinate Descent (RB-ICD) algorithm that updates subsets of phase-encoding lines simultaneously instead of one line at a time. This batched updating reduces optimization time, making scan-adaptive mask learning practical for cardiac cine series. Second, dSUNO extends the original test-time neighbor selection from static 2D images to a spatiotemporal neighbor search over entire cardiac cine series, enabling scan-adaptive mask estimation while maintaining temporal consistency across frames.

We further propose a novel model-based deep learning framework, termed \textbf{Mo}del-Based \textbf{S}patiotemporal Ne\textbf{t}work (\textbf{MostNet}), that integrates a MoDL~\cite{aggarwal2018modl}-style unrolled data-consistency architecture with CRNN-based spatiotemporal regularization modules inspired by CTFNet~\cite{qin2021complementary}.
MostNet uses both spatial and temporal correlations to achieve high-quality dynamic reconstructions under high undersampling factors.
Together, the proposed RB-ICD sampling technique, the test-time neighbor search over dynamic cine series for estimating the dSUNO masks, and the MostNet reconstruction network form a pipeline for efficient and high-quality dynamic cardiac MRI reconstruction.

The remainder of this paper is organized as follows.
Section~\ref{sec:theory} introduces the MRI acquisition model
and outlines our proposed subset-based randomized ICD optimization framework
for learning scan-adaptive Cartesian sampling patterns.
Section~\ref{sec:methods} describes the datasets, preprocessing pipeline,
and implementation details used for training and evaluation.
Section~\ref{sec:results} presents quantitative and qualitative results
on both the public OCMR dataset and our in-house MCardiac dataset.
Section~\ref{sec:discussion} summarizes our key findings
and discusses implications for clinical deployment.
Final conclusions are presented in Section~\ref{sec:conclusion}.

\section{Theory}\label{sec:theory}
 \subsection{MostNet-based Dynamic MRI Reconstruction}
\label{sec:mostnet_recon}

In cardiac cine MRI reconstruction, the task is to reconstruct a dynamic image series 
$\mathbf{x} \in \mathbb{C}^{n \times N_t}$ 
from undersampled multi-coil $k$-space measurements 
$\{\mathbf{y}_i \in \mathbb{C}^{m \times N_t}\}_{i=1}^{n_c}$. 
Here $n$ is the number of pixels in the spatial frame, $N_t$ is the number of temporal frames (cardiac phases), and $n_c$ is the number of coils. The forward model for each coil $i$ is given by:
\begin{equation}
\mathbf{y}_i = \mathbf{M} \mathbf{F} \mathbf{S}_i \mathbf{x},
\label{eq:forward_model}
\end{equation}
where $\mathbf{M}$ is the binary sampling mask, $\mathbf{F}$ is the 2D spatial Fourier transform applied independently to each frame, and $\mathbf{S}_i$ is the diagonal coil sensitivity matrix for coil $i$.

\begin{figure}
    \centering
    \includegraphics[width=\linewidth]{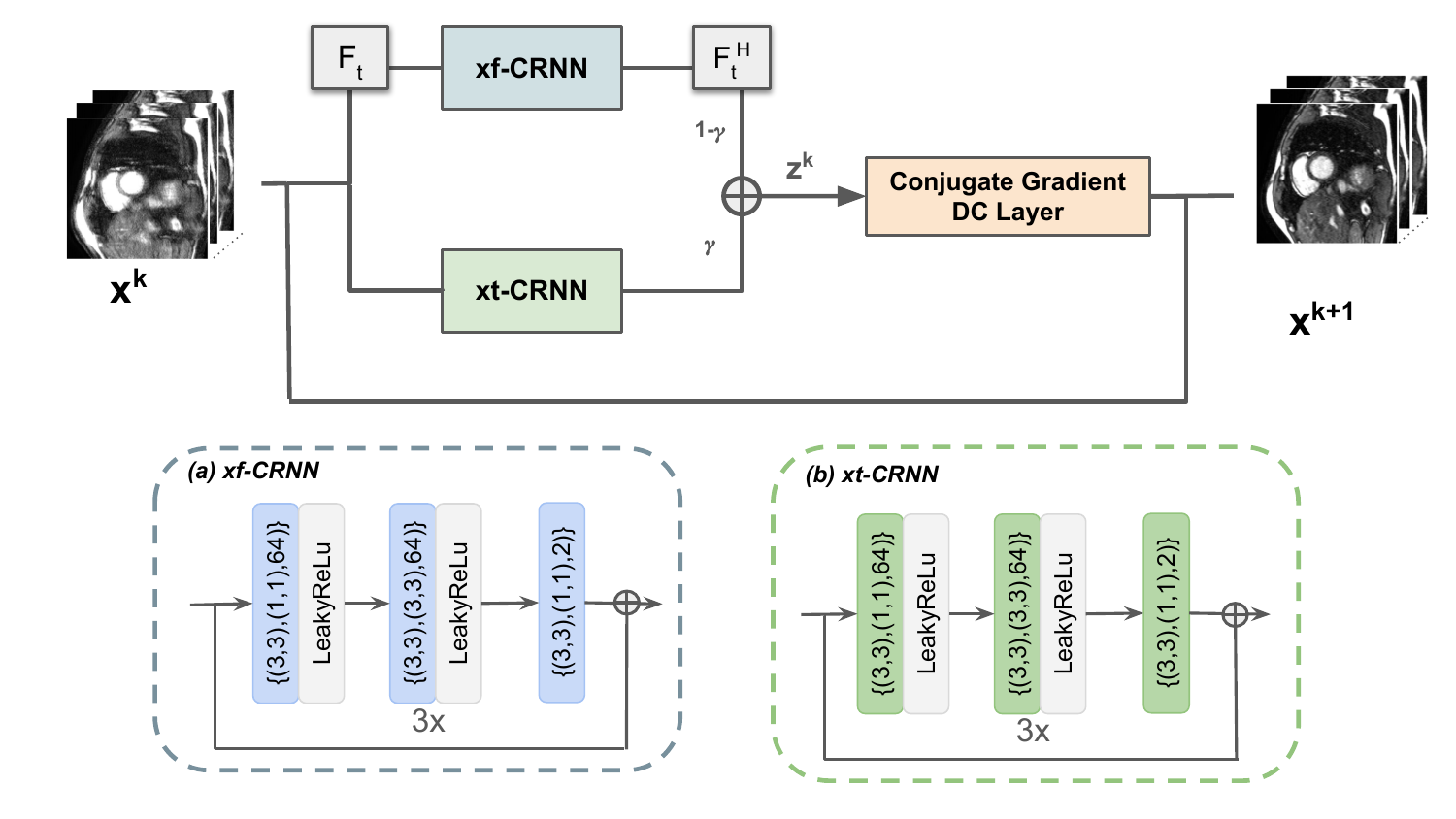}
    \caption{Schematic of the proposed dynamic MostNet reconstruction framework. At each iteration, a dual-domain CRNN denoiser operating in the spatiotemporal (x-t) and temporal-frequency (x-f) domains is followed by a conjugate-gradient data-consistency (DC) update. The CRNN blocks follow the complementary dual-domain design introduced in CTFNet~\cite{qin2021complementary}. The notation ``3x" indicates that the corresponding block is repeated
    three times within each denoising module.
    The learnable parameters $\T$ reside in the denoising blocks, while the DC layer
    enforces fidelity to the acquired multi-coil measurements.}
    \label{fig:MostNet_block_diagram}
\end{figure}

Motivated by the CTFNet dynamic reconstruction network~\cite{qin2021complementary}, which utilizes complementary temporal information via CRNN blocks, and the MoDL framework~\cite{aggarwal2018modl}, which enforces physics-based consistency via unrolled optimization, we propose a hybrid framework termed the \textbf{Mo}del-Based \textbf{S}patiotemporal Ne\textbf{t}work (\textbf{MostNet}). This approach integrates the dual-domain ($x-t$ and $x-f$) regularization of CTFNet with the conjugate-gradient data consistency mechanism of MoDL. An overview of the proposed framework is illustrated in Fig.~\ref{fig:MostNet_block_diagram}.

Specifically, we formulate the reconstruction as an optimization problem where the solution balances data fidelity with a learned spatiotemporal prior. The objective function is given by:
\begin{equation}
\arg\min_{\mathbf{x}} \;
\sum_{i=1}^{n_c} \left\| \mathbf{M} \mathbf{F} \mathbf{S}_i \mathbf{x} - \mathbf{y}_i \right\|_2^2
+ \lambda \left\| \mathbf{x} - \tilde{\mathcal{D}}_{\theta}(\mathbf{x}) \right\|_2^2,
\label{eq:objective}
\end{equation}
where $\lambda > 0$ is a regularization parameter controlling the trade-off between data fidelity and the prior. The term $\tilde{\mathcal{D}}_{\theta}$ represents the learned spatiotemporal denoising operator with parameters $\theta$, defined as:
\begin{equation}
\tilde{\mathcal{D}}_{\theta}(\mathbf{x})
= \gamma \cdot \mathcal{D}_{\mathrm{xt}}(\mathbf{x})
+ (1-\gamma) \cdot \mathcal{F}_t^{-1} \left( \mathcal{D}_{\mathrm{xf}} \left( \mathcal{F}_t\,\mathbf{x} \right) \right),
\label{eq:denoiser}
\end{equation}
where $\gamma \in [0,1]$ is a hyperparameter balancing the contribution of the two domains. $\mathcal{D}_{\mathrm{xt}}$ and $\mathcal{D}_{\mathrm{xf}}$ denote learned denoising operators acting in the spatiotemporal image ($x-t$) and temporal-frequency ($x-f$) domains, respectively, implemented using Convolutional Recurrent Neural Network (CRNN) blocks. Here, $\mathcal{F}_t$ denotes the Fourier transform applied along the temporal dimension.

To solve Eq.~\eqref{eq:objective}, we use an unrolled optimization scheme consisting of $K$ iterations (or stages). At each unrolled iteration $k$, we first apply the dual CRNN denoising operator to the current estimate $\mathbf{x}^k$:
\begin{equation}
\mathbf{z}^k = \tilde{\mathcal{D}}_{\theta}(\mathbf{x}^k),
\label{eq:denoise_update}
\end{equation}
followed by a data consistency (DC) update:
\begin{equation}
\mathbf{x}^{k+1} = \arg\min_{\mathbf{x}} \;
\sum_{i=1}^{n_c} \left\| \mathbf{M} \mathbf{F} \mathbf{S}_i \mathbf{x} - \mathbf{y}_i \right\|_2^2
+ \lambda \left\| \mathbf{x} - \mathbf{z}^k \right\|_2^2.
\label{eq:cg_update}
\end{equation}
We solve the quadratic subproblem in Eq.~\eqref{eq:cg_update} using a fixed number of Conjugate Gradient (CG) iterations. This process is repeated for $k = 0, \dots, K-1$, forming a $K$-stage unrolled network $f_{\theta}$ that jointly learns spatiotemporal priors and enforces physical measurement fidelity. 

Compared to CTFNet~\cite{qin2021complementary}, which relies on variable splitting and closed-form DC steps, MostNet adopts the complementary $x-t$ and $x-f$ CRNN blocks strictly as a learned regularization prior within a MoDL-style optimization. By explicitly solving the data-consistency problem via CG at each iteration, MostNet effectively decouples the learned prior from the physics-based reconstruction enforcement.

\subsection{Joint Optimization Framework for Scan-Adaptive Mask Learning}
\label{sec:joint_mask_recon_update}

This section details our proposed framework for jointly learning a set of scan-adaptive Cartesian undersampling patterns ($\mathbf{M}_i$) and a corresponding reconstruction network ($f_{\theta}$) specifically trained to perform well across these learned patterns.
Unlike the static imaging case where the underlying signal is a single 2D image, here we consider a dynamic image series $\mathbf{x} \in \mathbb{C}^{n \times N_t}$ consisting of $N_t$ temporal frames (cardiac phases), where $n$ is the number of pixels in each spatial frame. Using a training set of $N$ fully sampled dynamic $k$-space and corresponding ground truth image series, we define the joint optimization problem as:

\begin{equation}
\min_{\theta,\ \{ \mathbf{M}_i \}_{i=1}^N \subset \mathcal{C}}
\ \sum_{i=1}^N
\left\| f_{\theta}\!\left( \mathbf{A}_i^{H}\mathbf{M}_i \mathbf{y}^{\mathrm{full}}_i \right)
- \mathbf{x}^{\mathrm{gt}}_i \right\|_2^2
\label{eq:joint_opt}
\end{equation}

where $\mathbf{M}_i \in \mathcal{C}$ is the $i$-th training $k$-space subsampling mask chosen from the set $\mathcal{C}$ of all 1D Cartesian patterns with a specified sampling budget. The terms $\mathbf{y}^{\text{full}}_i$ and $\mathbf{x}^{\text{gt}}_i$ represent the $i$-th fully-sampled multi-coil training $k$-space and its corresponding ground truth cine series, respectively.

The operator $\mathbf{A}_i^H$ denotes the adjoint of the multi-coil spatiotemporal MRI forward operator for the $i$-th training cardiac cine series. Specifically, the forward operator $\mathbf{A}_i$ applies the coil sensitivity maps and 2D Fourier transform independently to each of the $N_t$ temporal frames, followed by multiplication with the corresponding sampling mask. Unacquired samples in $k$-space are zero-filled prior to applying the inverse Fourier transform. The term $f_{\theta}$ denotes the unrolled reconstruction network, whose learnable parameters $\theta$ correspond to the weights of the spatiotemporal denoising operator $\tilde{\mathcal{D}}_{\theta}$ used within each iteration of the unrolled reconstruction.

We solve this highly challenging, non-convex optimization problem using an alternating optimization framework. We alternate between updating the sampling masks $\{\mathbf{M}_i\}$ and updating the reconstruction network weights $\theta$. For a fixed reconstruction model, the scan-adaptive sampling masks are updated to minimize reconstruction error. However, given the high dimensionality of the dynamic cine series, the standard line-by-line update used in previous work is computationally prohibitive. To address this, we utilize the proposed Randomized Subset-based Iterative Coordinate Descent (RB-ICD) algorithm, described in the following subsection.

\subsection{Randomized Subset-based Iterative Coordinate Descent-based Mask Optimization (RB-ICD)}
\label{sec:rb_icd}

This section proposes a novel randomized subset-based ICD (RB-ICD) algorithm for optimizing scan-adaptive Cartesian sampling masks. This algorithm builds upon the iterative coordinate descent (ICD)–based mask optimization framework introduced in our prior work~\cite{gautam2026scan}, and extends it to dynamic cardiac cine MRI.

A key challenge in our previously proposed ICD-based mask optimization is computational scalability for dynamic cine data. In particular, updating one phase-encoding line at a time becomes computationally intensive because evaluating each candidate update requires reconstructing an entire multi-frame spatiotemporal cine series (rather than a single static image). Consequently, the per-update cost scales with the temporal length of the cine series in addition to the spatial dimensions. To address this, we introduce a randomized subset-based update technique, where multiple phase-encoding locations are updated simultaneously. This randomized batching reduces the number of expensive full-series reconstructions while maintaining sufficient exploration of the discrete mask search space.

The RB-ICD algorithm iteratively refines a sampling mask by relocating subsets of phase-encoding lines to previously non-sampled locations. Starting from an initial sampling mask, the algorithm proceeds in multiple passes over the mask ($N_{\text{iter}}$). Within each pass, subsets of size $s$ are selected from the currently sampled phase-encoding lines (excluding a fixed low-frequency ACS region).

For each subset, the selected $s$ phase-encoding lines are relocated to non-sampled locations to generate a set of $N_{\text{cand}}$ candidate masks.
Each candidate mask is evaluated by undersampling the dynamic cine series using that mask and computing the corresponding reconstruction error with a specified reconstruction model.
If a candidate mask yields an improvement over the current mask, the update is accepted.
This update procedure is repeated across subsets within a pass and over a fixed number of RB-ICD iterations $N_{\text{iter}}$ over the mask.
The overall procedure is summarized in Algorithm~\ref{alg:randomized_icd}.

\begin{algorithm}[!t]
\caption{RB-ICD Sampling Optimization}
\label{alg:randomized_icd}
\begin{algorithmic}[1]
\Require Fully sampled multi-coil $k$-space $\y$, ground-truth cine $\x_{\mathrm{gt}}$, initial mask $\m_{\mathrm{init}}$, reconstruction model $f$, loss $L$, number of phase encoding lines $N_y$, number of RB-ICD passes $N_{\mathrm{iter}}$, subset size $s$, number of candidate masks $N_{\mathrm{cand}}$, fixed ACS set $\Omega_{\mathrm{low}}$. 
\State $\m \leftarrow \m_{\mathrm{init}}$; enforce $\m(\Omega_{\mathrm{low}})=1$

\State $\mathcal{L}_{\mathrm{curr}} \leftarrow L(\x_{\mathrm{gt}},\, f(\m\y))$

\For{$j=1:N_{\mathrm{iter}}$} \Comment{passes}

    \State $\Omega_{\mathrm{mov}} \leftarrow \Omega(\m)\setminus \Omega_{\mathrm{low}}$ \Comment{$\Omega(\m)$: sampled locations of $\m$}


\State Randomly partition $\Omega_{\mathrm{mov}}$ (without replacement) into $N_{\mathrm{sub}}$ disjoint subsets $\{\Omega_{\mathrm{mov}}^{(t)}\}_{t=1}^{N_{\mathrm{sub}}}$, where $N_{\mathrm{sub}} = |\Omega_{\mathrm{mov}}|/s$.

    \For{$t=1:N_{\mathrm{sub}}$} \Comment{subset updates}
        \State $\Omega_{\mathrm{sub}} \leftarrow \Omega_{\mathrm{mov}}^{(t)}$
        
        \State $\Omega_{\mathrm{avail}} \leftarrow \{1,\dots,N_y\}\setminus \Omega(\m)$
        
        \State $\m_{\mathrm{best}} \leftarrow \m$; $\mathcal{L}_{\mathrm{best}} \leftarrow \mathcal{L}_{\mathrm{curr}}$
        
        \For{$p=1:N_{\mathrm{cand}}$} \Comment{candidate relocations}
        
            \State Sample $\Omega_{\mathrm{add}} \subset \Omega_{\mathrm{avail}}$ such that $|\Omega_{\mathrm{add}}| = s$.
            
            \State Form $\m_{\mathrm{cand}}$ from $\m$ by relocating $\Omega_{\mathrm{sub}}$ to $\Omega_{\mathrm{add}}$

            \State $\mathcal{L}_{\mathrm{cand}} \leftarrow L(\x_{\mathrm{gt}},\, f(\m_{\mathrm{cand}}\y))$
            
            \If{$\mathcal{L}_{\mathrm{cand}} < \mathcal{L}_{\mathrm{best}}$}
            
                \State $\mathcal{L}_{\mathrm{best}} \leftarrow \mathcal{L}_{\mathrm{cand}}$; $\m_{\mathrm{best}} \leftarrow \m_{\mathrm{cand}}$
                
            \EndIf
        \EndFor
        \If{$\mathcal{L}_{\mathrm{best}} < \mathcal{L}_{\mathrm{curr}}$}
            \State $\m \leftarrow \m_{\mathrm{best}}$
            \State $\mathcal{L}_{\mathrm{curr}} \leftarrow \mathcal{L}_{\mathrm{best}}$
        \EndIf
    \EndFor
\EndFor
\State \Return $\m$
\end{algorithmic}
\end{algorithm}

\subsection{Dynamic Scan-Adaptive Inference via Spatiotemporal Neighbor Search}
\label{sec:nn_search}

At test time, the objective is to select an optimized sampling pattern by identifying the training cardiac cine slice that most closely resembles the current test slice. To achieve this, we perform a spatiotemporal neighbor-based search that compares the test scan with a dictionary of pre-optimized training slices. Unlike the original static SUNO framework~\cite{gautam2026scan}, which identifies neighbors based solely on static 2D images, the proposed dynamic extension (dSUNO) performs neighbor search over entire cardiac cine series. Each candidate in the training dictionary corresponds to a full cine series with an associated optimized sampling mask obtained using the RB-ICD algorithm (Algorithm~\ref{alg:randomized_icd}).

In this dynamic setting, a ``neighbor" refers to an entire cardiac cine slice, which is the full spatiotemporal cine series, rather than an individual 2D frame. The comparison is based on spatial similarity between the initially acquired test frame and frames from the training sequences. At inference time, we restrict the comparison to the first test frame, since only low-frequency $k$-space data from the initial frame is assumed to be available at the start of a clinical scan.

Using the low-frequency reconstruction of the first test frame as a reference, we compare it against local temporal neighborhoods within each training slice. Each local temporal neighborhood consists of three consecutive frames $(i-1, i, i+1)$ in order to account for variations in cardiac phase. For each candidate training slice, we compute the average normalized root mean square difference (NRMSD) between the test frame and each such neighborhood:
\begin{equation}
    d_i = \frac{1}{3 \left\| \mathbf{x}_{\text{test}} \right\|_2} 
    \sum_{j=-1}^{1} 
    \left\| \mathbf{x}_{\text{test}} - \mathbf{x}^{\text{train}}_{i+j} \right\|_2,
    \label{eq:nrmse}
\end{equation}
where $\mathbf{x}_{\text{test}}$ denotes the low-frequency reconstruction of the first test frame and $\mathbf{x}^{\text{train}}_{i+j}$ denotes the low-frequency reconstruction of the training slice at temporal offset $j$ relative to frame index $i$. The index $i$ ranges from $2$ to $N_t-1$, and for each training slice, the value of $i$ that minimizes $d_i$ is selected. The nearest neighbor is finally selected as the training slice that yields the lowest NRMSD across the entire training set.

The optimized sampling mask associated with this nearest-neighbor slice is then assigned to the current test scan, producing the proposed dSUNO mask. This technique enables robust, scan-adaptive mask prediction using only the spatial information available at the beginning of the scan. 

\section{Methods}\label{sec:methods}

\begin{figure}[!ht]
    \centering
    \includegraphics[width=0.95\linewidth]{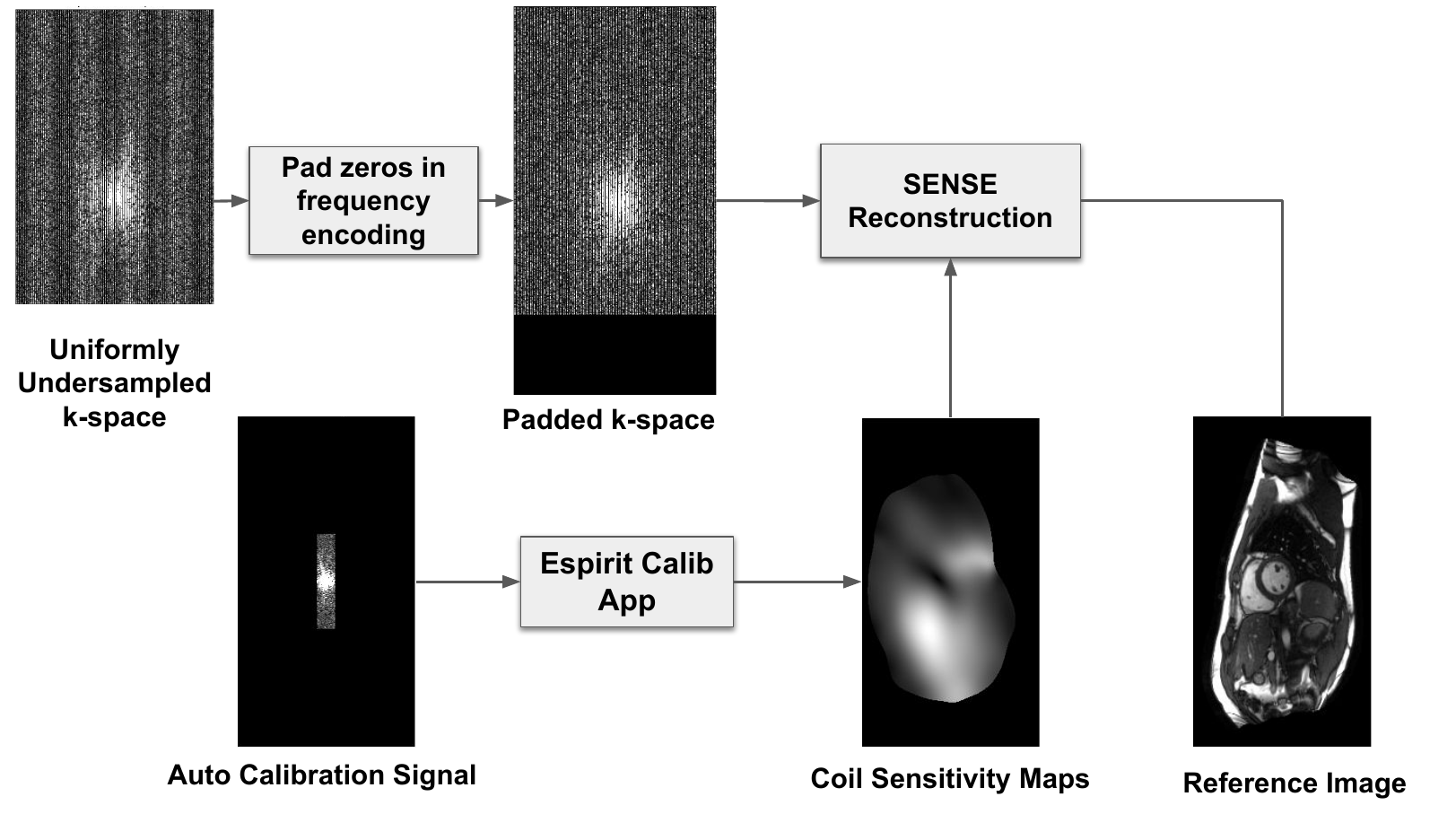}
    \caption{Ground truth reconstruction pipeline for the MCardiac dataset. ESPIRiT~\cite{uecker2014espirit}-based coil sensitivity maps are computed from the ACS region, and SENSE reconstruction is performed on zero-padded $k$-space to obtain reference images.}
    \label{fig:ucardiac_gt_pipeline}
\end{figure}

\subsection{Datasets}
To evaluate the proposed RB-ICD sampling method, we conducted experiments on two cardiac cine datasets: the publicly available OCMR dataset and an in-house clinical dataset. These datasets allow us to assess performance across both fully sampled and partially sampled acquisition settings.

\subsubsection{MCardiac Dataset}
\label{subsec:ucardiac_data}

The in-house dataset (referred to as {MCardiac}) was acquired at the University of Michigan using a 1.5T Sola MRI scanner (Siemens Healthineers, Erlangen, Germany). Data were acquired with a 30-channel receiver coil array using a standard 2D Cartesian cine MRI protocol (balanced SSFP pulse sequence, TR = 3.02 ms, flip angle = 59$\degree$, and a field-of-view ranging from 300–350 mm in both the readout and phase-encoding directions). Each dynamic sequence consists of 20–25 temporal frames with a temporal resolution of approximately 30–35 ms per frame, acquired across 10–12 heartbeats. Whole left ventricular coverage was achieved by acquiring 12–16 slices with a thickness of 8 mm in the short-axis orientation.

\begin{figure}[!ht]
    \centering
    \includegraphics[width=0.7\linewidth]{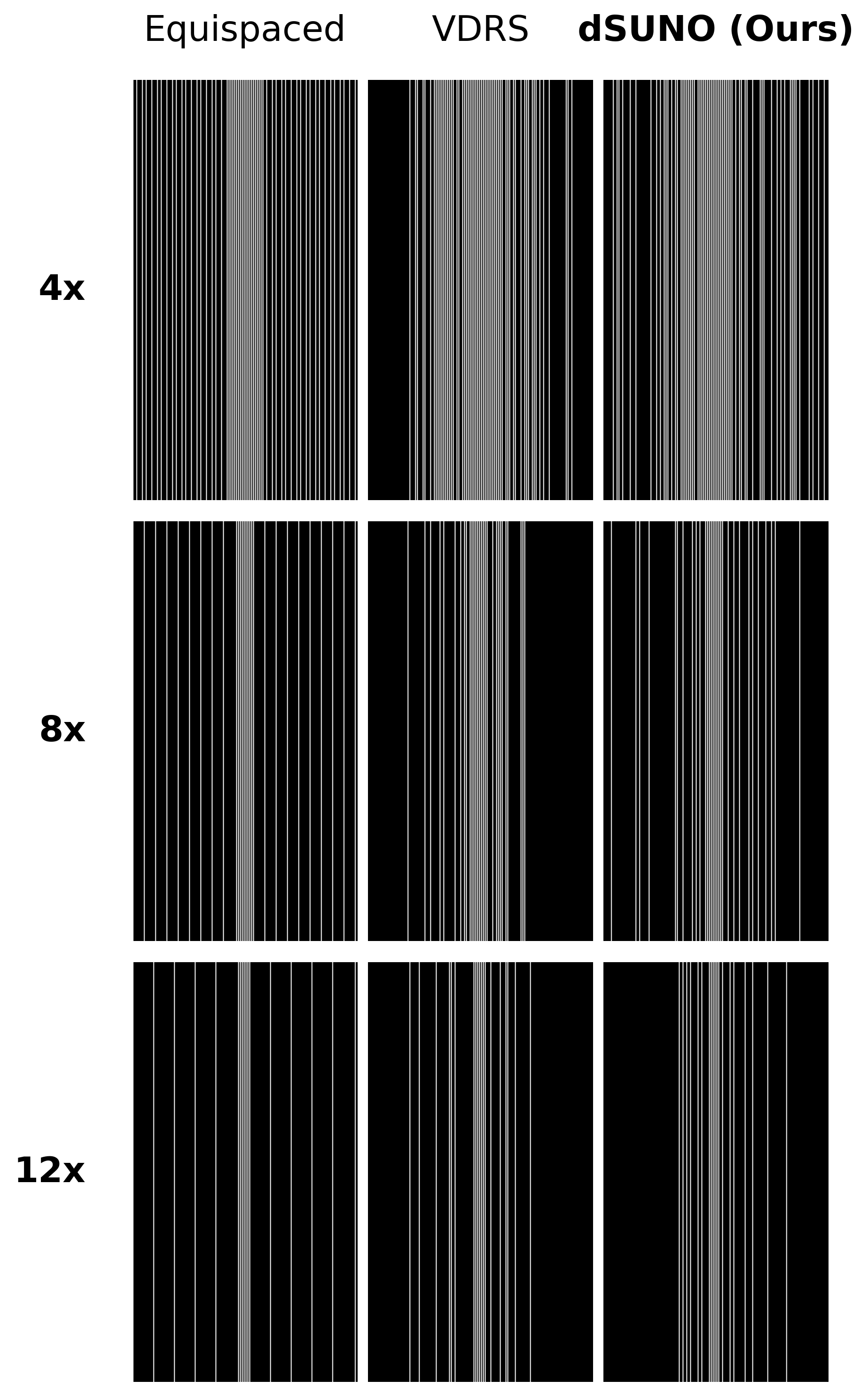}
    \caption{Comparison of Cartesian undersampling masks for the MCardiac dataset. The grid displays two baseline sampling masks - Equispaced (left) and VDRS (middle) alongside the proposed dSUNO masks (right) at 4$\times$, 8$\times$, and 12$\times$ acceleration factors. These represent the retrospective undersampling patterns selected from the originally acquired prospective 2$\times$ grid.}
    \label{fig:baseline_masks}
\end{figure}

The native acquisition matrix was $440 \times 240$
in the readout and phase-encoding directions.
All acquisitions included a fully sampled central auto-calibration signal (ACS) region.
Data were originally acquired using a uniform 2$\times$ undersampling pattern
in the phase-encoding direction. To simulate higher acceleration rates (4$\times$, 8$\times$, and 12$\times$), we applied additional retrospective undersampling masks directly to the originally acquired 2$\times$ undersampled $k$-space. All reconstruction experiments were performed on this doubly undersampled data.

To generate ground truth images for evaluation, we performed SENSE reconstruction using coil sensitivity maps estimated from the fully sampled ACS region. 
Fig.~\ref{fig:ucardiac_gt_pipeline} illustrates
the full preprocessing and ground truth generation pipeline.
The sensitivity maps were estimated using ESPIRIT~\cite{pruessmann1999sense, uecker2014espirit}.
The dataset comprised 14 subjects in total, split at the subject level
into 10 training subjects and 4 test subjects to avoid overlap.
Each subject contributed 13 short-axis slices,
resulting in 130 training slices and 52 test slices,
with each slice treated as a cardiac cine sequence.

\subsubsection{OCMR Dataset}

The OCMR dataset~\cite{chen2020ocmr} contains multi-coil cine cardiac MRI data acquired on Siemens platforms (Prisma 3T, Avanto 1.5T, and Sola 1.5T), including both fully sampled and prospectively undersampled scans. We focused on the short-axis (sax) cine scans from the fully sampled subset acquired at 1.5T using an 18-channel coil. Each sequence contains 20–30 temporal frames.
For this dataset, the data was split
into 139 training and 26 testing slices.
Retrospective undersampling masks were applied to simulate acceleration factors of 4$\times$ and 6$\times$. This dataset served as a benchmark for evaluating undersampling and reconstruction strategies.

\subsection{Mask Optimization}
\label{sec:mask_opt}

We followed the alternating minimization framework described in Section~\ref{sec:joint_mask_recon_update} to optimize scan-adaptive undersampling masks for both the MCardiac and the OCMR datasets.
We initialized the process using the variable-density random sampling (VDRS) masks and first trained a U-Net reconstruction model on images undersampled using these VDRS masks. This trained U-Net was then used as the reconstruction model within the RB-ICD framework (Algorithm~\ref{alg:randomized_icd}) to obtain an initial set of scan-adaptive sampling masks, which were optimized independently for each dynamic cine slice and shared across all temporal frames within a slice. 

The scan-adaptive RB-ICD masks obtained from this stage were then used to train a two-channel MoDL reconstruction network.
Using this trained MoDL as the reconstruction model inside RB-ICD, we
performed a second round of mask optimization to obtain refined scan-adaptive masks. Finally, both MoDL, CTFNet, and the proposed MostNet models were trained on the training images undersampled by these optimized scan-adaptive masks.

For mask optimization, we used the parameters summarized in Table~\ref{tab:ucardic_icd_param}.
For 4$\times$ acceleration for the MCardiac dataset, each mask had a total sampling budget of $B=60$ phase-encoding lines, including $F=20$ fixed low-frequency lines and $40$ movable lines, with a subset size of $s=10$ and three ICD
passes. For higher acceleration factors, the total sampling budget, number of movable lines, subset size, and number of ICD passes were adjusted accordingly, as listed in Table~\ref{tab:ucardic_icd_param}. A detailed ablation study analyzing the impact of the subset size $s$ and the number of iterations $N_{iter}$ on the RB-ICD algorithm's accuracy and runtime is provided in the Supplementary document.

\begin{table}[ht]
\centering
\begin{tabular}{lccc}
\toprule
\textbf{Parameter} & $\mathbf{4\times}$ & $\mathbf{8\times}$ & $\mathbf{12\times}$ \\
\midrule
Total budget ($B$)                 & 60    & 30    & 20    \\
Low-frequency lines ($F$)         & 20    & 10    & 6     \\
Movable lines ($B - F$)            & 40    & 20    & 14    \\
Subset size ($s$)                  & 10    & 5     & 3     \\
ICD Passes ($N_{iter}$)                    & 3     & 6     & 9     \\
\bottomrule
\end{tabular}
\caption{
RB-ICD optimization parameters for the MCardiac dataset (440 $\times$ 240) along the phase-encoding direction at acceleration factors 4$\times$, 8$\times$, and 12$\times$.}
\label{tab:ucardic_icd_param}
\end{table}


\subsection{Reconstruction Network Training}
\label{sec:recon_training}

We trained both MoDL and MostNet models on the MCardiac and OCMR datasets using the optimized scan-adaptive ICD masks at 4$\times$, 8$\times$, and 12$\times$ acceleration. Each dynamic scan was split into training and validation sets.
For MoDL~\cite{aggarwal2018modl}, we used a UNet~\cite{ronneberger2015u} as the denoiser inside the training framework, in which six unrollings of the denoiser and the conjugate gradient (CG) blocks were used.
The regularization parameter $\lambda$ controlling the weighting of the two terms in MoDL was set to $10^{-2}$, and the tolerance for the CG algorithm used was $10^{-5}$.
For MostNet, we used six cascades of unrolled architecture using the CRNN-based spatiotemporal regularization.
The relative weighting of the CRNN prior and the regularization parameter was set
to $\gamma = 0.5$ and $\lambda=10^{-2}$, respectively.
The network was trained using an $\ell_2$ loss and the Adam optimizer with a learning rate of $10^{-4}$ for 100 epochs. Sensitivity analyses justifying the choice of the regularization parameter and the number of unrolled iterations are provided in the Supplementary document.

\subsection{Baseline Sampling Masks}
\label{sec:baseline_masks}

We compared our dSUNO optimized masks against two commonly used sampling techniques:
variable density random sampling (VDRS)~\cite{lustig2007sparse}
and equispaced sampling~\cite{haldar2010compressed}. For experiments on the public OCMR dataset, which provides fully sampled $k$-space data,
we additionally include comparisons with the VISTA spatiotemporal sampling
method~\cite{ahmad2015variable}.
In VDRS, one-third of the budgeted lines were fixed at the center of $k$-space,
and the rest were sampled randomly using a polynomial-decay distribution~\cite{lustig2007sparse}.
For equispaced masks, one-third of the lines were again fixed centrally,
with the remainder spaced uniformly outside the center.
All masks were created separately for each dataset and acceleration factor.
For MCardiac, masks were applied on top of the existing 2$\times$ uniform undersampling.
Figure~\ref{fig:baseline_masks} shows representative examples.

\subsection{Evaluation Metrics}
\label{sec:evaluation_metrics}
We evaluated reconstruction quality using image similarity metrics, qualitative comparisons, and clinical measures.
We report normalized mean square error (NMSE), structural similarity index measure (SSIM)~\cite{wang2004image}, and peak signal-to-noise ratio (PSNR):
\begin{equation}
    \text{NMSE}(\x,\hat{\x}) = \frac{\|\x-\hat{\x}\|_2^2}{\|\x\|_2^2},
\end{equation}
\begin{equation}
\text{SSIM}(\x,\hat{\x}) =
\frac{(2\mu_{\x}\mu_{\hat{\x}} + c_1)(2\sigma_{\x\hat{\x}} + c_2)}
{(\mu_{\x}^2 + \mu_{\hat{\x}}^2 + c_1)(\sigma_{\x}^2 + \sigma_{\hat{\x}}^2 + c_2)},
\end{equation}
\begin{equation}
\text{PSNR}(\x,\hat{\x}) =
10 \log_{10}\!\left(
\frac{\max(|\x|)^2\, d}{\|\x - \hat{\x}\|_2^2}
\right),
\end{equation}
where $\x$ denotes the ground-truth image,
$\hat{\x}$ denotes the reconstructed image,
$d$ is the total number of pixels,
$\mu_{\x}$ and $\mu_{\hat{\x}}$ are the mean intensities,
$\sigma_{\x}^2$ and $\sigma_{\hat{\x}}^2$ are the variances,
and $\sigma_{\x\hat{\x}}$ is the covariance between $\x$ and $\hat{\x}$.
The constants $c_1$ and $c_2$ are small positive stabilization constants
used to avoid numerical instability when the denominators are close to zero.
 



\section{Results}\label{sec:results}
\begin{figure}[t]
    \centering
    \includegraphics[width=0.9\textwidth]{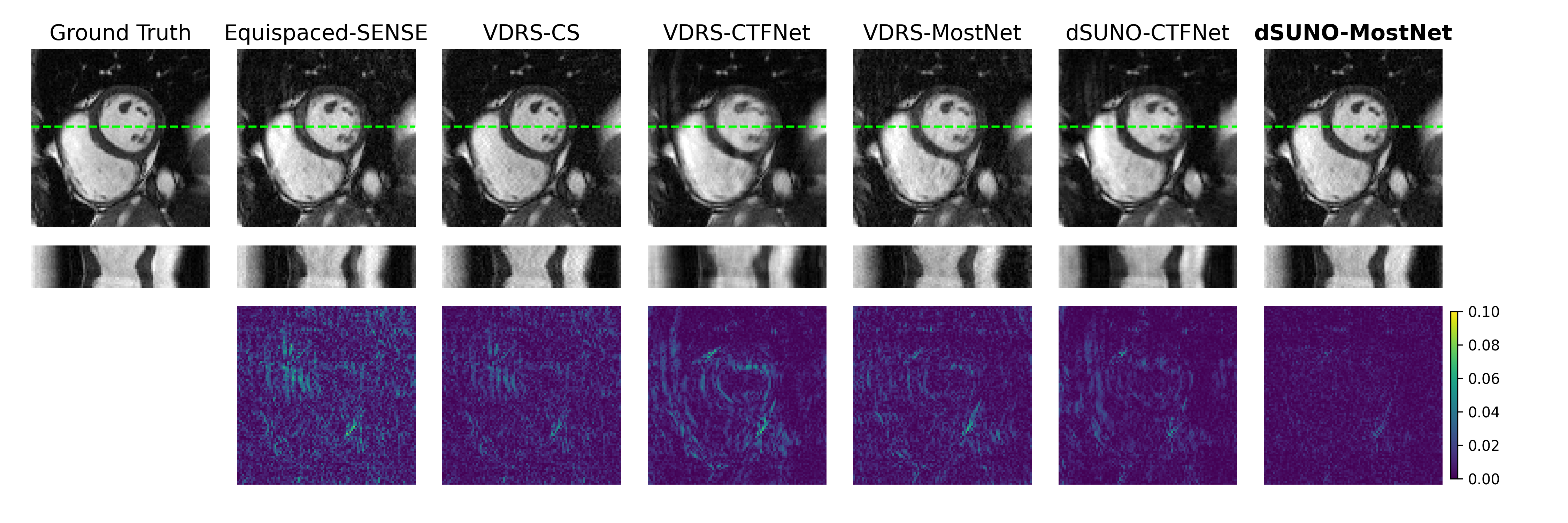}
    
    \includegraphics[width=0.9\textwidth]{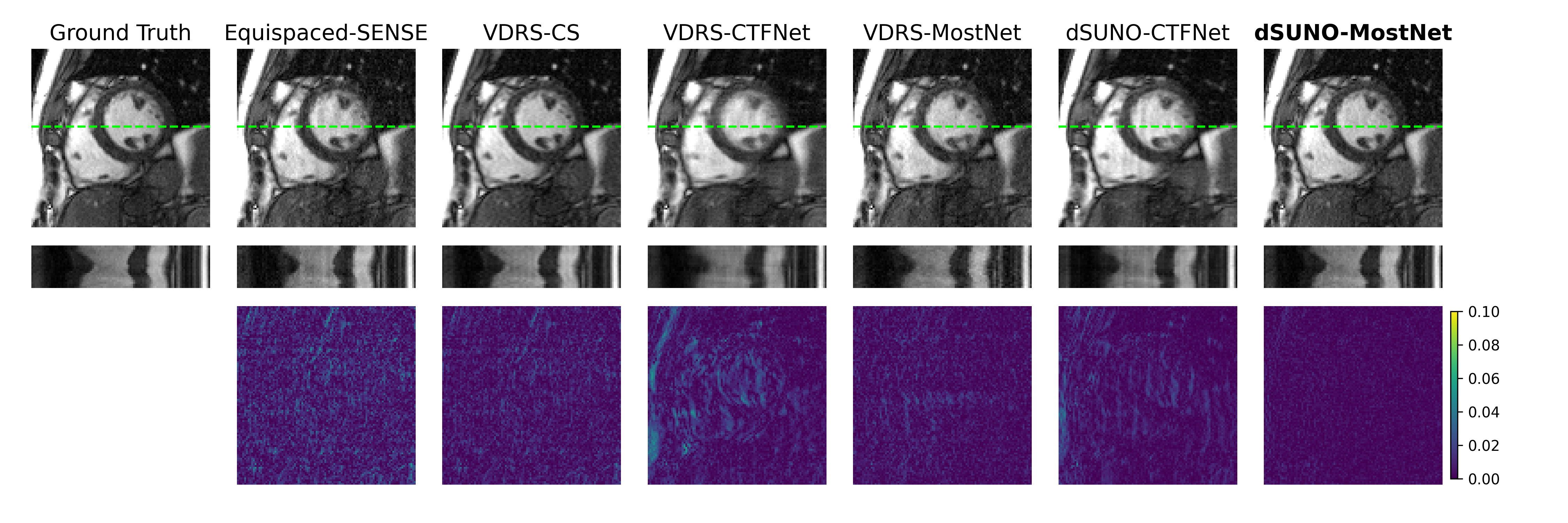}
    \caption{
    ROI-based qualitative comparison of different sampling techniques and reconstructions on two anatomically distinct slices at 4$\times$ acceleration. 
    Each figure includes: (i) reconstructed image cropped over the region-of-interest (ROI) around the heart, (ii) spatio–temporal (x-t) profile from a horizontal stripe within the ROI, and (iii) error maps showing the absolute difference from the ground truth over the ROI. The dSUNO–MostNet reconstructions show improved artifact suppression and temporal stability relative to the baselines.}
    \label{fig:qualitative_xt_error_4x_multi}
\end{figure}

\begin{figure}[t]
    \centering
    \includegraphics[width=0.9\textwidth]{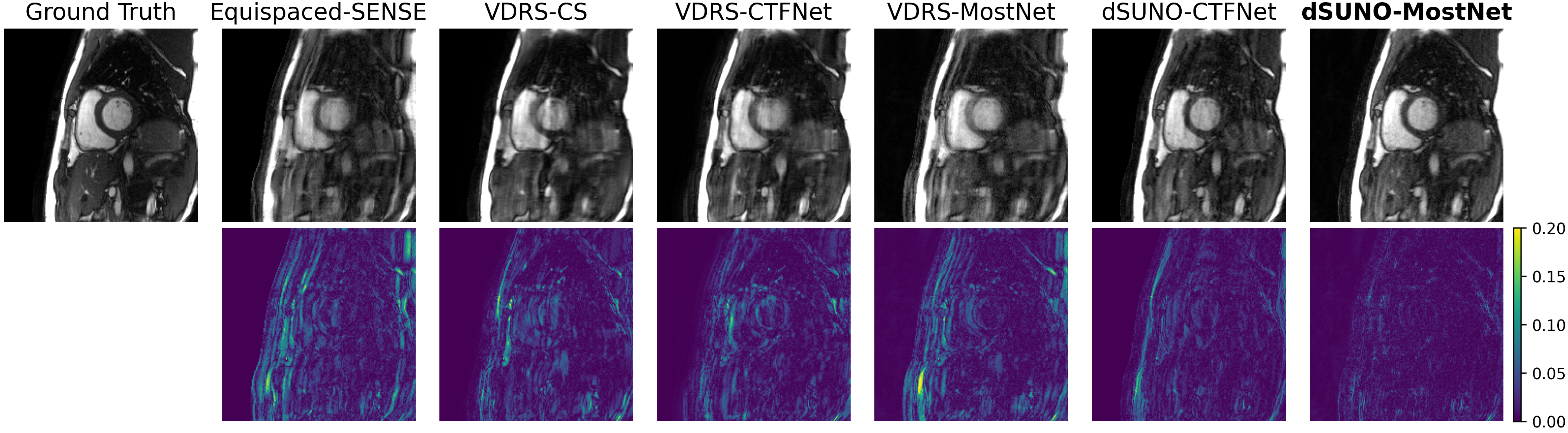}
    \includegraphics[width=0.9\textwidth]{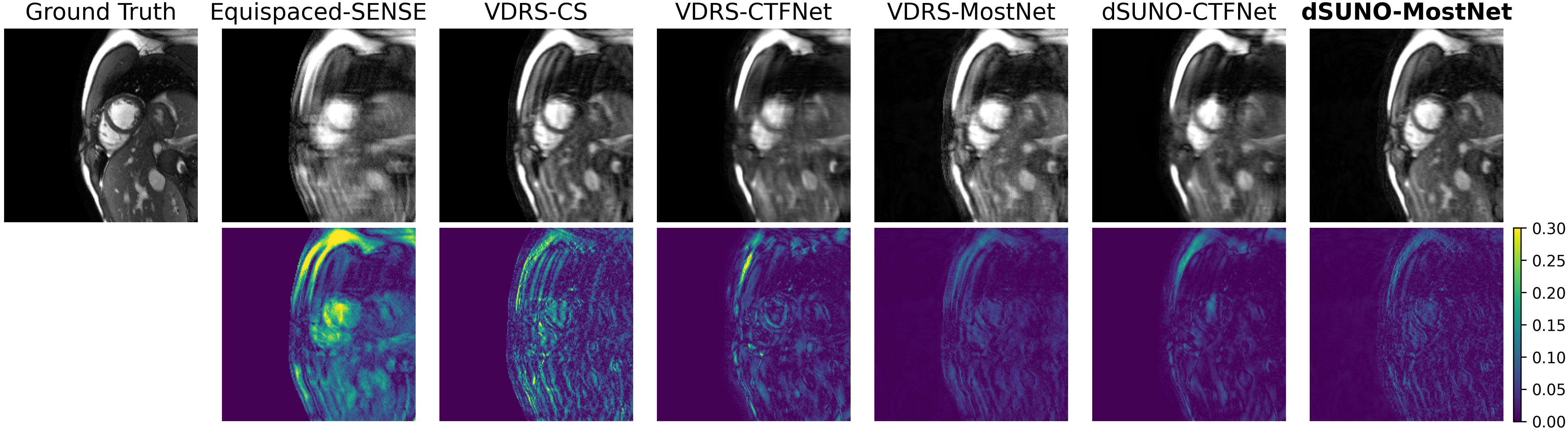}
    \caption{
Qualitative comparison of dSUNO with other baseline sampling techniques at 8$\times$ (top panel) and 12$\times$ (bottom panel) acceleration factors for two different slices.
Each panel includes the (i) reconstructed images and (ii) error maps computed against ground truth for the same frame.
As the undersampling becomes more aggressive, reconstructions using the equispaced and VDRS mask degrade noticeably, while dSUNO-MostNet continues to preserve anatomical details.}
 \label{fig:qualitative_xt_error_8x_12x}
\end{figure}

\subsection{Evaluation on the MCardiac Dataset}\label{sec:mcardiac_eval}
The proposed sampling and reconstruction framework was evaluated on the in-house MCardiac dataset at acceleration factors of $4\times$, $8\times$, and $12\times$.
For each training slice, an optimized undersampling mask was generated using the RB-ICD algorithm (Section~\ref{sec:rb_icd}).
During testing, for each slice, the corresponding dSUNO mask was obtained from the pre-optimized slice-adaptive masks using the nearest-neighbor search as described in Section~\ref{sec:nn_search}.

For our experiments, dSUNO masks were compared with the equispaced~\cite{haldar2010compressed} and VDRS~\cite{lustig2007sparse} masks, commonly used for cardiac cine MRI.
Equispaced masks were reconstructed using SENSE~\cite{pruessmann1999sense} while the VDRS masks were reconstructed using traditional compressed sensing (CS)~\cite{lustig2007sparse} as well as learned reconstruction models, including the CTFNet and the proposed MostNet.
Similarly, dSUNO masks were also tested using the same choice of dynamic reconstruction models, CTFNet, and the proposed MostNet model.
Each learned reconstruction model was trained separately for the sampling patterns with which it was evaluated.

The commonly used dynamic sampling baseline, VISTA~\cite{ahmad2015variable}, was not included in the MCardiac experiments because the data were prospectively acquired using a fixed uniform $2\times$ Cartesian undersampling pattern in the phase-encoding direction.
As VISTA is a prospective spatiotemporal acquisition technique, its application would require access to fully sampled $k$-space data, which is not available for this dataset.

Figure~\ref{fig:qualitative_xt_error_4x_multi} shows qualitative reconstructions at 4$\times$ acceleration for two representative test slices, comparing dSUNO masks with commonly used baseline sampling techniques, each paired with a reconstruction method typically used with that sampling approach.
The top row of the figure shows the ground truth along with the reconstructed images using a particular combination of mask and reconstruction model, the middle row shows the corresponding spatio–temporal ($x$–$t$) profiles of the images, and the bottom row shows the absolute error maps.
Reconstructions obtained using dSUNO masks with MostNet exhibit reduced aliasing and improved temporal coherence relative to baseline techniques.
These qualitative comparisons reflect differences in both sampling pattern and reconstruction model.
Figure~\ref{fig:qualitative_xt_error_8x_12x} presents the results for $8\times$ and $12\times$ acceleration factors, where differences between the proposed dSUNO and the rest of the baselines across methods become more pronounced. Even under highly accelerated conditions, reconstructions using dSUNO masks better preserve the endocardial border and papillary muscle visibility.

Table~\ref{tab:sampling_comparison} summarizes the quantitative results over all evaluated test slices and shows that the proposed dSUNO sampling consistently improves reconstruction quality relative to equispaced and VDRS baselines. In particular, the dSUNO–MostNet configuration achieves the lowest NMSE and highest PSNR at all acceleration factors. 
While dSUNO-CTFNet attains the highest SSIM at $4\times$ and $8\times$, dSUNO-MostNet achieves the highest SSIM at $12\times$ and consistently yields the lowest NMSE and highest PSNR across all acceleration factors.

The performance gap between dSUNO-based methods and fixed-mask baselines becomes more pronounced at higher acceleration factors, where conventional approaches degrade substantially. These results demonstrate that the proposed dSUNO technique enables robust and effective reconstruction for highly accelerated dynamic cardiac MRI when combined with state-of-the-art reconstruction models.

\begin{table}[t]
\centering
\begin{tabular}{clcccccc}
\toprule
\textbf{$AF$} & \textbf{Metric}
& \shortstack{Equispaced\\SENSE}
& \shortstack{VDRS\\CS}
& \shortstack{VDRS\\CTFNet}
& \shortstack{VDRS\\MostNet}
& \shortstack{dSUNO\\CTFNet}
& \shortstack{\textbf{dSUNO}\\\textbf{MostNet}} \\
\midrule
\multirow{3}{*}{4}
& NMSE      
& 0.051 & 0.055 & 0.047 & 0.042 & 0.039 & \textbf{0.024} \\
& PSNR
& 34.78 & 34.49 & 34.91 & 35.03 & 35.65 & \textbf{37.99} \\
& SSIM      
& 0.962 & 0.948 & 0.921 & 0.939 & \textbf{0.964} & 0.960 \\
\midrule
\multirow{3}{*}{8}
& NMSE      
& 0.190 & 0.227 & 0.155 & 0.152 & 0.147 & \textbf{0.086} \\
& PSNR 
& 28.14 & 27.61 & 29.05 & 29.12 & 30.23 & \textbf{32.31} \\
& SSIM      
& 0.863 & 0.857 & 0.826 & 0.844 & \textbf{0.898} & 0.892 \\
\midrule
\multirow{3}{*}{12}
& NMSE      
& 0.455 & 0.408 & 0.251 & 0.240 & 0.254 & \textbf{0.114} \\
& PSNR
& 24.49 & 25.14 & 26.78 & 27.00 & 27.48 & \textbf{30.85} \\
& SSIM      
& 0.789 & 0.792 & 0.801 & 0.800 & 0.851 & \textbf{0.875} \\
\bottomrule
\end{tabular}
\caption{Quantitative comparison on the MCardiac test set at acceleration factors $AF=4\times$, $8\times$, and $12\times$.
The proposed dSUNO consistently improves reconstruction quality over other baselines with both the CTFNet and MostNet architectures. dSUNO-MostNet achieves the lowest NMSE and highest PSNR across all acceleration factors.
For SSIM, dSUNO–CTFNet achieves the highest values at 4$\times$ and 8$\times$, while dSUNO–MostNet performs best at 12$\times$.
}
\label{tab:sampling_comparison}
\end{table}


\subsection{Evaluation on the OCMR Dataset}
In this section, we evaluate the proposed dSUNO framework on the publicly available multicoil OCMR dataset~\cite{chen2020ocmr} and compare it with commonly used baseline sampling techniques. Unlike the in-house MCardiac dataset, OCMR provides fully sampled dynamic cardiac cine scans, enabling retrospective undersampling. The dataset consists of 2D short-axis balanced SSFP acquisitions across multiple subjects and cardiac phases.

For training, scan-adaptive dSUNO masks were optimized using the proposed RB-ICD algorithm (Algorithm~\ref{alg:randomized_icd}). During testing, the corresponding dSUNO mask was estimated using the nearest neighbor search (described in Section~\ref{sec:nn_search}). We compared dSUNO against widely used baseline sampling techniques, including equispaced Cartesian sampling, VDRS, and the spatiotemporal VISTA scheme~\cite{ahmad2015variable}. Specifically, equispaced masks were reconstructed using SENSE~\cite{pruessmann1999sense}, VDRS masks were reconstructed using compressed sensing~\cite{lustig2007sparse} as well as learned reconstruction models, including CTFNet and the proposed MostNet. VISTA masks were also reconstructed using both the CTFNet~\cite{qin2021complementary} and the proposed MostNet reconstruction network.

Table~\ref{tab:ocmr_sampling_comparison} summarizes quantitative results in terms of NMSE, PSNR, and SSIM. Across all acceleration factors, dSUNO masks consistently improve reconstruction accuracy relative to fixed-mask baselines. In particular, the proposed dSUNO-MostNet configuration achieves the lowest NMSE and highest PSNR at all acceleration factors, while the VISTA-CTFNet baseline attains the highest SSIM.

At higher accelerations, the advantages of scan-adaptive sampling become more pronounced. Particularly, at $6\times$ and $8\times$, dSUNO-MostNet substantially improves reconstruction quality compared to VDRS and VISTA, indicating improved robustness under aggressive undersampling. Overall, these results demonstrate that the proposed dSUNO framework provides consistent performance gains on OCMR and that combining scan-adaptive dSUNO sampling with the proposed MostNet reconstruction yields improved reconstruction quality under high acceleration.

\begin{table}[t]
\centering
\begin{tabular}{clccccccccc}
\toprule
\textbf{$AF$} & \textbf{Metric} 
& \shortstack{Equispaced\\SENSE} 
& \shortstack{VDRS\\CS} 
& \shortstack{VDRS\\CTFNet} 
& \shortstack{VDRS\\MostNet}
& \shortstack{VISTA\\CTFNet} 
& \shortstack{VISTA\\MostNet}
& \shortstack{dSUNO\\CTFNet}
& \shortstack{\textbf{dSUNO}\\\textbf{MostNet}} \\
\midrule

\multirow{3}{*}{4}
& NMSE      
& 0.122 & 0.116 & 0.110 & 0.080 & 0.112 & 0.070 & 0.075 & \textbf{0.046} \\
& PSNR (dB)
& 27.85 & 28.14 & 28.33 & 29.72 & 28.29 & 30.45 & 29.91 & \textbf{32.22} \\
& SSIM      
& 0.864 & 0.813 & 0.815 & 0.824 & 0.870 & 0.872 & \textbf{0.875} & 0.865 \\

\midrule
\multirow{3}{*}{6}
& NMSE      
& 0.224 & 0.188 & 0.174 & 0.163 & 0.151 & 0.128 & 0.144 & \textbf{0.083} \\ 
& PSNR (dB)
& 25.17 & 25.67 & 25.82 & 26.59 & 26.99 & 27.90 & 27.08 & \textbf{29.74} \\
& SSIM      
& 0.741 & 0.770 & 0.738 & 0.751 & 0.805 & 0.812 & 0.801 & \textbf{0.818} \\

\midrule
\multirow{3}{*}{8}
& NMSE      
& 0.492 & 0.305 & 0.220 & 0.211 & 0.163 & 0.158 & 0.191 & \textbf{0.103} \\
& PSNR (dB)
& 21.89 & 24.23 & 25.12 & 25.43 & 26.64 & 26.98 & 25.83 & \textbf{27.71} \\
& SSIM      
& 0.704 & 0.711 & 0.702 & 0.715 & 0.761 & 0.768 & 0.759 & \textbf{0.773} \\
\bottomrule
\end{tabular}
\caption{Quantitative comparison of reconstruction quality on the OCMR test set at acceleration factors 
$AF = 4\times$, $6\times$, and $8\times$.  
Metrics include normalized mean square error (NMSE), peak signal-to-noise ratio (PSNR), and structural similarity index (SSIM). 
The proposed dSUNO-MostNet framework achieves the lowest NMSE and highest PSNR across all acceleration factors. dSUNO-CTFNet attains the highest SSIM at 4$\times$ while dSUNO-MostNet achieves the highest SSIM at 6$\times$ and 8$\times$.}
\label{tab:ocmr_sampling_comparison}
\end{table}


\subsection{Effect of Dynamic Reconstruction Model}
\label{sec:recon_progression_mcardiac}
In this section, we demonstrate how explicitly modeling temporal dynamics in the reconstruction leads to improved reconstruction quality for dynamic cardiac MRI compared to static, frame-wise reconstruction. To this end, we compare the static MoDL reconstruction model~\cite{aggarwal2018modl}, which processes each temporal frame independently, with the proposed dynamic MostNet reconstruction model, which jointly models temporal correlations across frames. 

To enable a fair comparison, we consider three configurations: VDRS-MoDL, dSUNO-MoDL, and dSUNO-MostNet. 
This comparison enables us to first assess the benefit of replacing conventional fixed VDRS sampling with the proposed scan-adaptive dSUNO under a static reconstruction model, and then to isolate the additional gains obtained by replacing the static reconstruction model with the proposed dynamic reconstruction model.

As shown in Table~\ref{tab:recon_progression}, dSUNO yields a substantial improvement over VDRS  even when using the static MoDL reconstruction. More importantly, using the same dSUNO masks, the proposed dynamic MostNet consistently improves reconstruction quality over MoDL in terms of NMSE and PSNR across all acceleration factors. These results indicate that explicitly modeling temporal correlations in the reconstruction provides a clear advantage over static, frame-wise reconstruction approaches for dynamic cardiac MRI. 
While dSUNO–MoDL achieves the highest SSIM, dSUNO–MostNet attains lower reconstruction error (NMSE) and higher PSNR, reflecting improved overall reconstruction performance, particularly at higher acceleration factors.

\begin{table}[t]
\centering
\begin{tabular}{clccc}
\toprule
\textbf{$AF$} & \textbf{Metric}
& \shortstack{VDRS\\MoDL}
& \shortstack{dSUNO\\MoDL}
& \shortstack{\textbf{dSUNO}\\\textbf{MostNet}} \\
\midrule
\multirow{3}{*}{4}
& NMSE      & 0.049 & 0.031 & \textbf{0.024} \\
& PSNR (dB) & 34.91 & 37.24 & \textbf{37.99} \\
& SSIM      & 0.936 & \textbf{0.968} & 0.960 \\
\midrule
\multirow{3}{*}{8}
& NMSE      & 0.203 & 0.094 & \textbf{0.086} \\
& PSNR (dB) & 28.50 & {32.27} & \textbf{32.31} \\
& SSIM      & 0.847 & \textbf{0.932} & 0.892 \\
\midrule
\multirow{3}{*}{12}
& NMSE      & 1.221 & 0.128 & \textbf{0.114} \\
& PSNR (dB) & 20.08 & 30.54 & \textbf{30.85} \\
& SSIM      & 0.522 & \textbf{0.906} & 0.875 \\
\bottomrule
\end{tabular}
\caption{Comparison of the proposed dynamic MostNet reconstruction model with the static MoDL model on the MCardiac test set. The VDRS-MoDL configuration is included to assess the effect of the proposed dSUNO sampling using the static model. dSUNO-MostNet achieves the lowest NMSE and highest PSNR across all acceleration factors, while dSUNO-MoDL attains the highest SSIM.}

\label{tab:recon_progression}
\end{table}


\subsection{Qualitative Review by Radiologists}

We performed a blinded reader study
to assess the diagnostic quality and clinical interpretability
of reconstructed cardiac cine MR images.
Two cardiovascular radiologists independently evaluated short-axis cine slices
reconstructed from data acquired using the baseline undersampling patterns
(VDRS and equispaced)
as well as the proposed dSUNO-optimized sampling masks,
combined with both classical and deep learning–based dynamic reconstruction methods.

Reconstructed cine video in GIF format from multiple slices and subjects were intensity-normalized and presented in randomized order for 8$\times$ acceleration factors. Readers were blinded to both the sampling and the reconstruction method, and all cases were reviewed under identical display and playback conditions to ensure consistent assessment.

Each reconstruction was evaluated using a 5-point Likert scale (1 = non-diagnostic, 2 = poor, 3 = acceptable, 4 = good, 5 = excellent) across six diagnostic criteria:
(i) endocardial border sharpness,
(ii) blood–myocardium contrast,
(iii) temporal dynamics of the left-ventricular (LV) myocardium,
(iv) visibility of papillary muscles,
(v) severity of residual undersampling artifacts,
and (vi) overall diagnostic confidence.
Ratings were averaged across readers, slices, and subjects
for each sampling–reconstruction combination and acceleration factor.

\begin{figure}[t]
    \centering
    \includegraphics[width=1\linewidth]{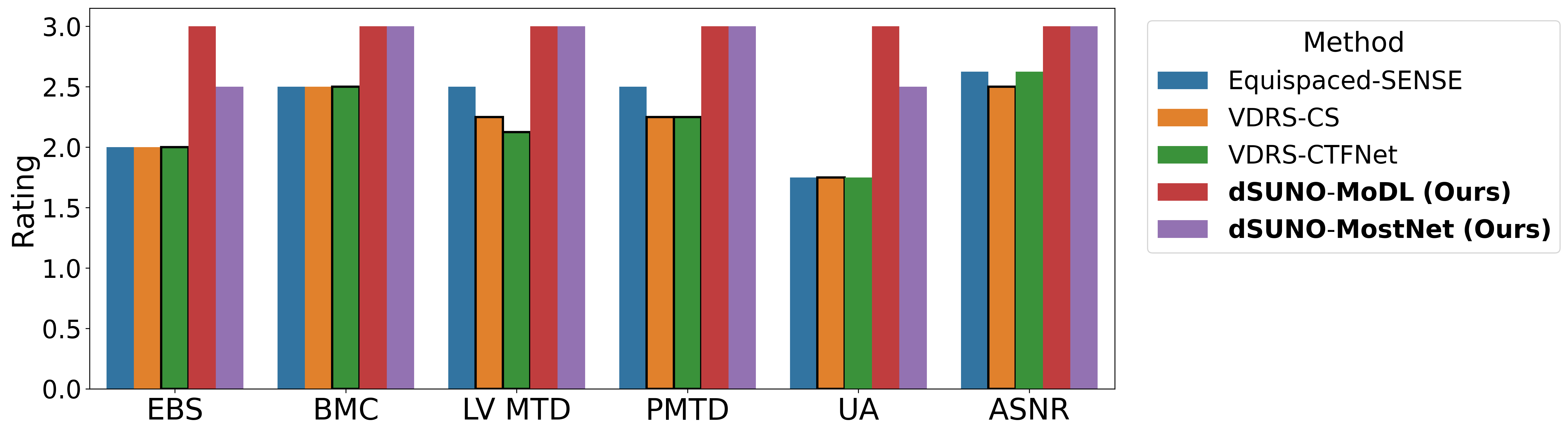}
    \caption{Blinded radiologist ratings of diagnostic quality at $8\times$ undersampling for different methods.
    Bars indicate mean Likert scores (1 = non-diagnostic, 5 = excellent) across six qualitative criteria. For compact visualization, the six criteria are abbreviated as EBS: endocardial border sharpness; BMC: blood-myocardium contrast;
    LV MTD: left-ventricular myocardial temporal dynamics;
    PMTD: papillary muscle temporal dynamics;
    UA: undersampling artifacts; ASNR: apparent signal-to-noise ratio.}
    \label{fig:rating_boxplot_8x}
\end{figure}

Figure~\ref{fig:rating_boxplot_8x} summarizes the average radiologist ratings at $8\times$ undersampling. Reconstructions obtained using dSUNO sampling with MoDL and MostNet consistently achieved the highest diagnostic scores across all categories, outperforming those using equispaced and VDRS sampling.
These perceptual findings align closely with the quantitative improvements
in Table~\ref{tab:sampling_comparison},
confirming that dSUNO sampling preserves clinically interpretable image quality
even under high acceleration.

\section{Discussion}\label{sec:discussion}
Compared to the prior SUNO framework~\cite{gautam2026scan} for static MRI, this work introduces a faster and more efficient scan-adaptive sampling technique for dynamic imaging. The previously proposed ICD mask optimization algorithm~\cite{gautam2026scan}, which updates one phase-encoding line at a time, is replaced with the proposed RB-ICD method that updates subsets of phase-encoding lines per iteration. This modification substantially accelerates mask optimization while maintaining competitive reconstruction quality in our experiments. Although developed for dynamic cine MRI here, RB-ICD is a general mask optimization routine and can be similarly applied to static MRI (e.g., knee or brain MRI) by evaluating candidate masks using a static reconstruction network.

Clinically, prospective scan-time savings from an acceleration factor are expected to scale approximately with the phase-encoding sampling budget; however, the exact reduction depends on the specific cine series implementation and gating-related overheads. In this study, the higher acceleration factors for the in-house MCardiac data are evaluated retrospectively, i.e., the data were acquired with a uniform prospective $2\times$ pattern and additional undersampling masks were applied offline to simulate $4\times/8\times/12\times$. Accordingly, we focus on image quality at matched sampling budgets rather than reporting absolute breath-hold time reductions.

Our ablation studies further show that a moderate subset size ($k$=4,8) provides the best trade-off between reconstruction accuracy and computational speed. Additionally, while standard designs like VISTA offer population-level solutions, our results demonstrate that scan-adaptive masks better preserve fine anatomical details, such as endocardial border sharpness, under high acceleration.

We emphasize that the proposed approach is slice-adaptive, assigning a single sampling mask to an entire cardiac cine slice. While frame-adaptive time-varying sampling could, in principle, better capture temporal variations, it is difficult to implement in routine cardiac cine MRI. In ECG-gated acquisition, $k$-space for each cardiac phase is accumulated over multiple heartbeats, and adapting the sampling pattern on a frame-by-frame basis would require updating patterns 
after partial data from each phase, along with repeated gating and retrospective binning. This can complicate the acquisition workflow and is not readily compatible with standard cine protocols or realistic scan-time constraints. The slice-adaptive formulation, therefore, represents a simplified yet deployable design choice for dynamic cardiac imaging.

\section{Conclusion}\label{sec:conclusion}
This work presented a scan-adaptive Cartesian undersampling framework for dynamic cardiac MRI, in which sampling masks are optimized offline using an alternating sampler–reconstructor training technique and selected at test time via a neighbor search. Mask optimization is performed using a novel randomized batched iterative coordinate descent (RB-ICD) algorithm. The proposed MostNet architecture integrates data-consistency layers from MoDL with temporal regularization modules inspired by CTFNet.
Experimental results on both the public OCMR dataset and the in-house MCardiac dataset show that the proposed scan-adaptive sampling technique consistently improves reconstruction quality over commonly used Cartesian sampling baselines across multiple acceleration factors. Compared to existing baseline sampling patterns, the proposed technique reduces aliasing artifacts and better preserves temporal consistency across frames. Radiologist evaluations further indicate that scan-adaptive masks lead to improved diagnostic image quality under aggressive undersampling.
The results suggest that tailoring Cartesian sampling patterns to individual dynamic cardiac MRI scans can provide benefits over fixed sampling patterns. Future work will investigate extensions to non-Cartesian and physics-informed sampling trajectories, as well as closer integration with self-supervised and deep image prior–based reconstruction methods.


\section*{Acknowledgment}
The authors gratefully acknowledge Prof. Maryam Sayadi (Michigan State University) for helpful discussions. The authors also acknowledge Evan Bell (currently at Johns Hopkins University) for contributions to this work while he was at Michigan State University.


\section*{Data Availability}
Code will be made publicly available upon acceptance at:
\url{https://github.com/sidgautam95/rbicd-dynamic-mri-sampling}.


\bibliographystyle{unsrt}
\bibliography{references}



\clearpage
\appendix

\section*{Supplementary Material}
\addcontentsline{toc}{section}{Supplementary Material}

\setcounter{section}{0}
\renewcommand{\thesection}{S\arabic{section}}

\setcounter{subsection}{0}
\renewcommand{\thesubsection}{S\arabic{section}.\arabic{subsection}}

\setcounter{figure}{0}
\renewcommand{\thefigure}{S\arabic{figure}}

\setcounter{table}{0}
\renewcommand{\thetable}{S\arabic{table}}


\section{Introduction}
This supplementary document provides the ablation studies for the Randomized Batched Iterative Coordinate Descent (RB-ICD) parameters and sensitivity analyses for the proposed MostNet reconstruction network hyperparameters used for our experiments.
This information justifies the parameter selections used in the main manuscript by characterizing the performance of the proposed RB-ICD algorithm as a function of subset size ($s$) and iteration count ($N_{iter}$), while also analyzing the sensitivity of MostNet reconstruction quality to changes in the regularization parameter ($\lambda$) and the number of unrolling iterations ($K$). All supplemental experiments were done on the MCardiac dataset at 4$\times$ acceleration.

\section{Ablation Studies on RB-ICD Sampling Parameters}
To understand the choices that affect the performance of our proposed randomized batched iterative coordinate descent (RB-ICD) sampling framework, we conducted some ablation studies. These analyses quantify the influence of key parameters on the learned sampling patterns and the resulting image reconstruction quality. Specifically, we varied the subset size $s$, the number of RB-ICD iterations $N_{\text{iter}}$, the loss function used for mask optimization, and the reconstruction network guiding the updates. Unless otherwise stated, all ablations were performed on the {MCardiac} dataset at an acceleration factor of 8$\times$, using MoDL-based reconstructions.

\subsection{Subset Size $s$}
\begin{figure}[ht]
\centering
\includegraphics[width=0.9\linewidth]{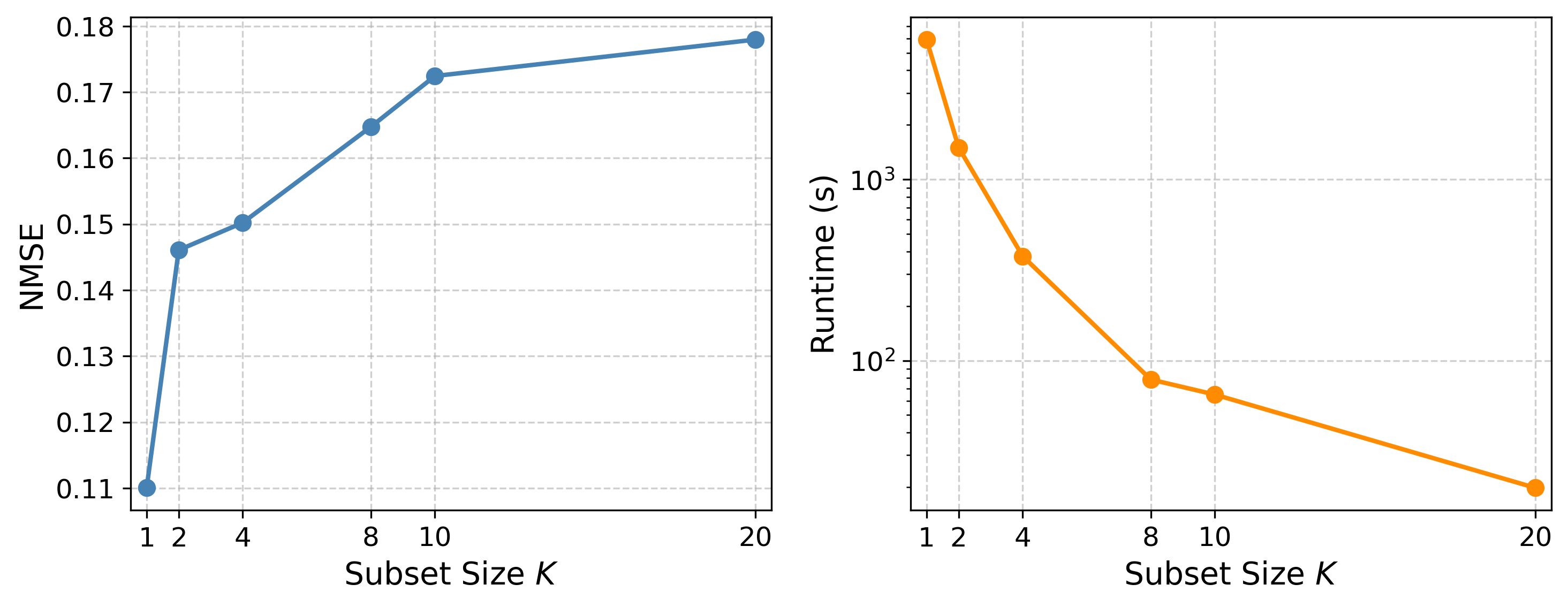}
\caption{\,Impact of subset size $s$ on reconstruction error and computational efficiency. Left: The reconstruction error (NMSE) increases with $s$, as larger subsets result in a less exhaustive exploration of the discrete search space. 
Right: Runtime decreases sharply and follows a logarithmic trend as $s$ increases, driven by the reduction in the number of disjoint subsets $N_{sub}$.}
\label{fig:abl_k_combined}
\end{figure}

The subset size $s$ controls the number of $k$-space lines updated per iteration within the RB-ICD sampling process. Smaller values of $s$ (e.g., $s=1$) enable fine-grained and localized mask refinements, generally yielding lower reconstruction error but at the expense of substantially higher runtime. 
Conversely, larger subset sizes increase the number of possible candidate updates; however, only $N_{\text{cand}}=20$ randomly sampled candidate masks are evaluated at each iteration. As a result, the optimization becomes more approximate for larger subset sizes, leading to a trade-off between computational efficiency and optimization accuracy.
We evaluated $s \in \{1, 2, 4, 8, 10, 20\}$ on the MCardiac dataset at 4$\times$ acceleration using MoDL-based reconstructions.
Figure~\ref{fig:abl_k_combined} shows that the reconstruction error (NMSE) gradually increases with $s$, as larger subset sizes lead to less exhaustive exploration of the discrete mask search space.
This indicates a clear trade-off between optimization accuracy and computational efficiency.
The runtime analysis shows that computational cost decreases sharply as $s$ increases because a larger subset size reduces the number of disjoint subsets $N_{sub}$, thereby requiring fewer expensive full-series reconstructions to evaluate the $N_{cand}$ candidate masks. 

\subsection{Number of ICD Iterations $N_{\text{iter}}$}

\begin{figure}[ht]
\centering
\includegraphics[width=0.9\linewidth]{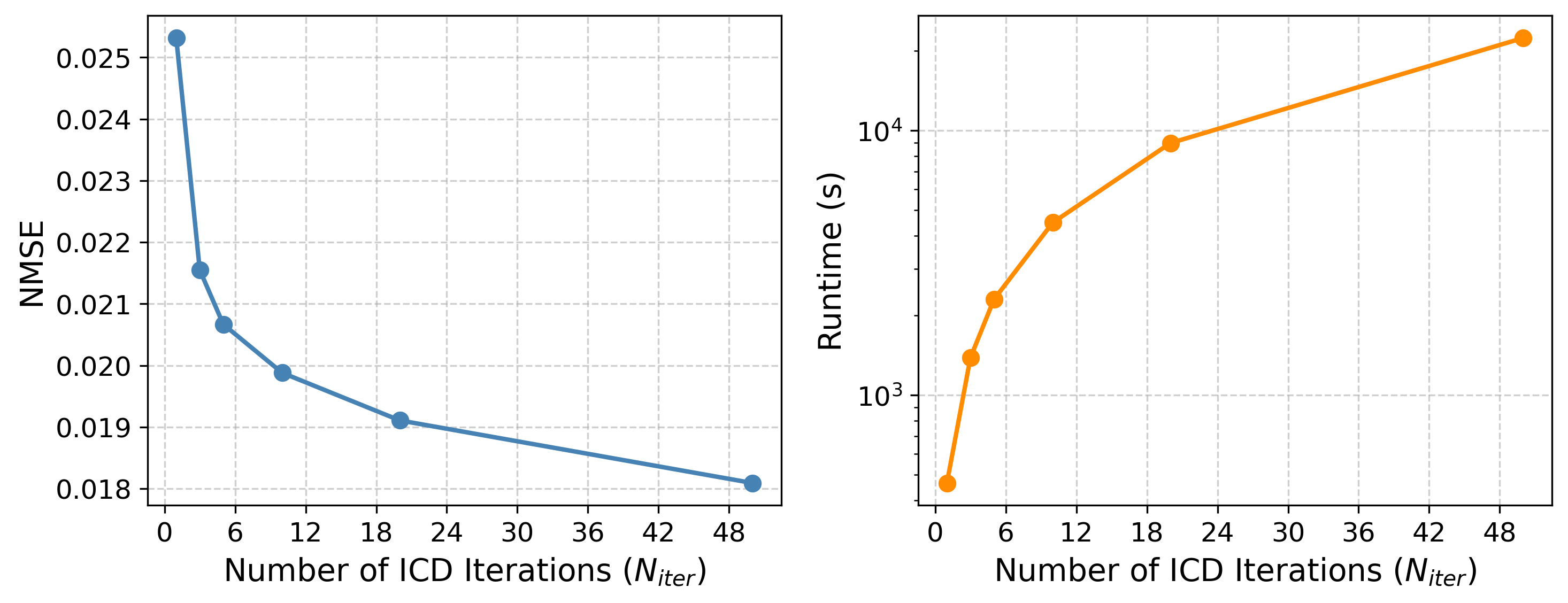}
\caption{\,Impact of the number of RB-ICD iterations $N_{\text{iter}}$ on reconstruction quality and computational overhead. Left: The reconstruction error (NMSE) decreases rapidly as $N_{\text{iter}}$ increases from 1 to 3, with diminishing gains in mask refinement observed beyond 5 iterations. Right: Runtime increases approximately linearly with $N_{\text{iter}}$, reflecting the predictable computational cost added by each successive round of subset-wise updates. 
}
\label{fig:abl_iter}
\end{figure}

The number of RB-ICD iterations ($N_{\text{iter}}$) determines how many rounds of subset-wise updates are performed to refine the sampling mask. Increasing $N_{\text{iter}}$ generally improves mask quality but also increases computation time.
We evaluated the impact of $N_{\text{iter}} \in \{1, 3, 5, 10, 20, 50\}$ on the MCardiac dataset using MoDL reconstruction model at 8$\times$ acceleration factor. As shown in the left panel of Figure~\ref{fig:abl_iter}, the Normalized Mean Square Error (NMSE) drops sharply as $N_{\text{iter}}$ increases from 1 to 3. Beyond 5 iterations, the reconstruction performance begins to plateau, yielding diminishing returns in mask refinement. The corresponding runtime analysis in the right panel shows that the computational overhead increases approximately linearly with $N_{\text{iter}}$. This linear trend reflects the predictable cost added by each successive round of subset-wise updates. Thus, 3–5 iterations offer the most efficient trade-off between optimization quality and practical computational efficiency.

\section{{MostNet Hyperparameter Sensitivity Analysis}}
In this section, we analyze the sensitivity of the proposed MostNet reconstruction network's performance to two key hyperparameters: the regularization parameter $\lambda$ and the number of unrolled iterations $K$. Experiments were performed using the MCardiac dataset at 4$\times$ acceleration, with performance evaluated using Peak Signal-to-Noise Ratio (PSNR) and Structural Similarity Index Measure (SSIM) reconstruction metrics. 
We use a univariate ablation protocol: one parameter is varied at a time while all other training and reconstruction settings are held fixed. All ablation results in this section were generated using the proposed dSUNO masks at 4$\times$ acceleration factor. For all ablation studies, the parameter $\gamma$ controlling the weighting between the xf-CRNN and xt-CRNN blocks was fixed at 0.5.

\subsection{Regularization Parameter ($\lambda$)}
\begin{figure}[t]
    \centering
    \begin{subfigure}[b]{0.48\linewidth}
        \centering
        \includegraphics[width=\linewidth]{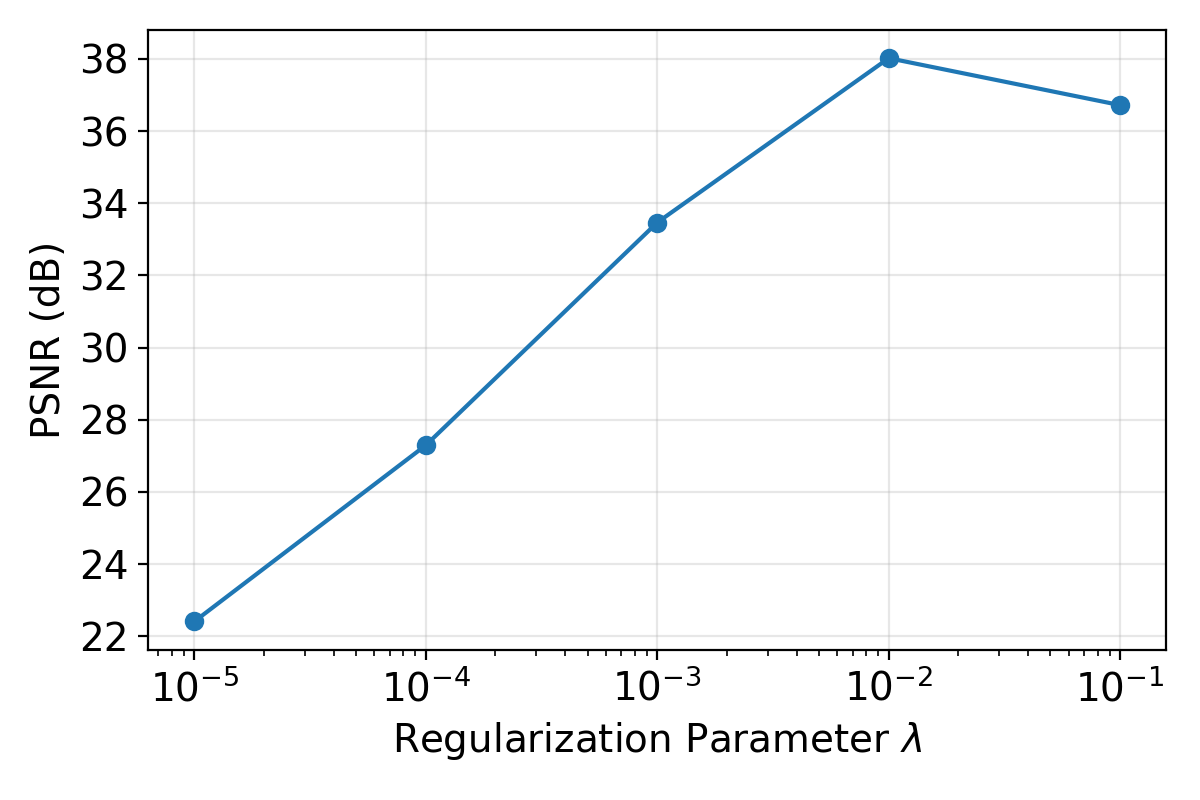}
    \end{subfigure}
    \hfill
    \begin{subfigure}[b]{0.48\linewidth}
        \centering
        \includegraphics[width=\linewidth]{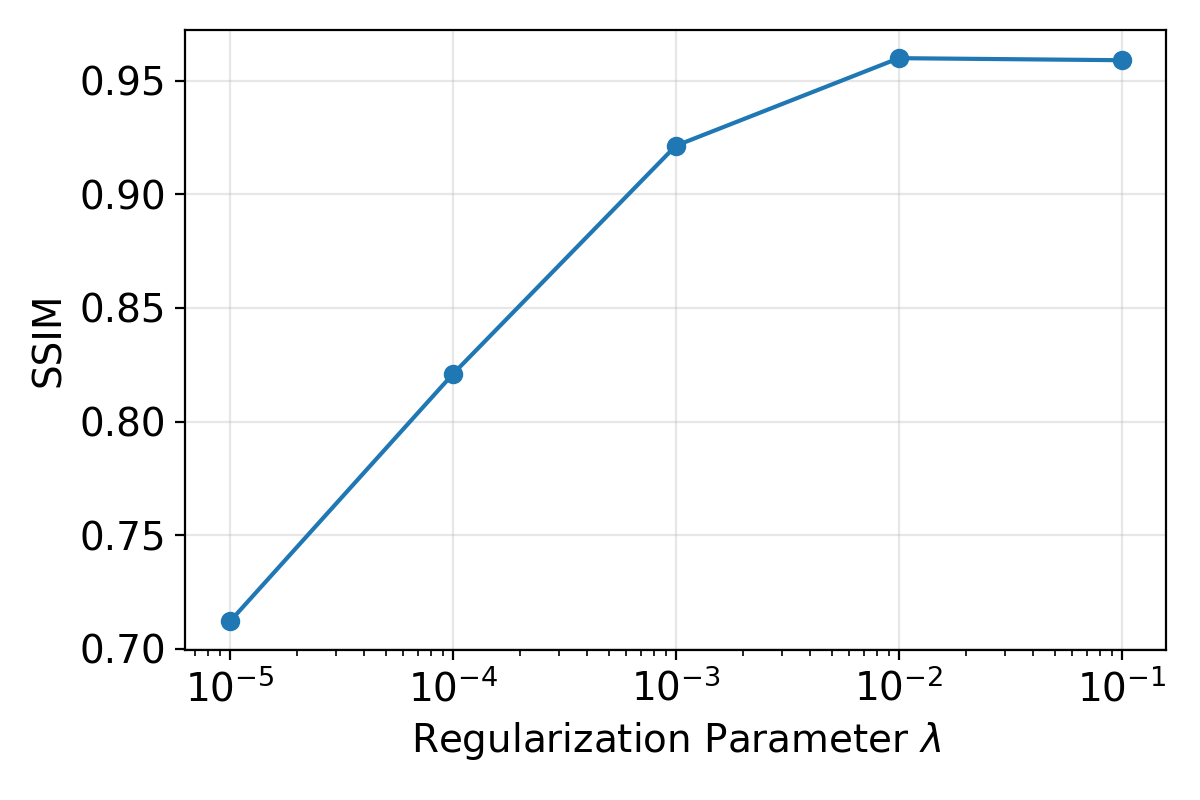}
    \end{subfigure}
    \caption{\, Sensitivity of MostNet reconstruction quality to the regularization parameter $\lambda$ with $K=6$ fixed. $\lambda=10^{-2}$ gives the best reconstruction quality in terms of PSNR and SSIM among the values.}
    \label{fig:ablation_lambda_both}
\end{figure}

The parameter $\lambda$ in Eq.~(2) of the main manuscript governs the trade-off between the data-fidelity term and the learned prior $\|\x-\tilde{\mathcal{D}}(\x)\|_2^2$. Figure~\ref{fig:ablation_lambda_both} shows the effect of varying $\lambda$ on reconstruction performance on the test set (with $K=6$ and $\gamma=0.5$ fixed). Performance improves as $\lambda$ increases from small values, reaches an optimum at $\lambda = 10^{-2}$, and then degrades for larger values. For low values ($\lambda < 10^{-3}$), the data-fidelity term dominates, leading to insufficient denoising and residual aliasing artifacts. For high values ($\lambda > 10^{-1}$), the learned prior over-regularizes the solution, yielding over-smoothed images and loss of fine detail. Overall, these trends confirm that $\lambda= 10^{-2}$ offers the best performance in terms of PSNR and SSIM for the reconstructed images.

\subsection{Number of Unrolled Iterations ($K$)}
\begin{figure}[t]
    \centering
    \begin{subfigure}[b]{0.48\linewidth}
        \centering
        \includegraphics[width=\linewidth]{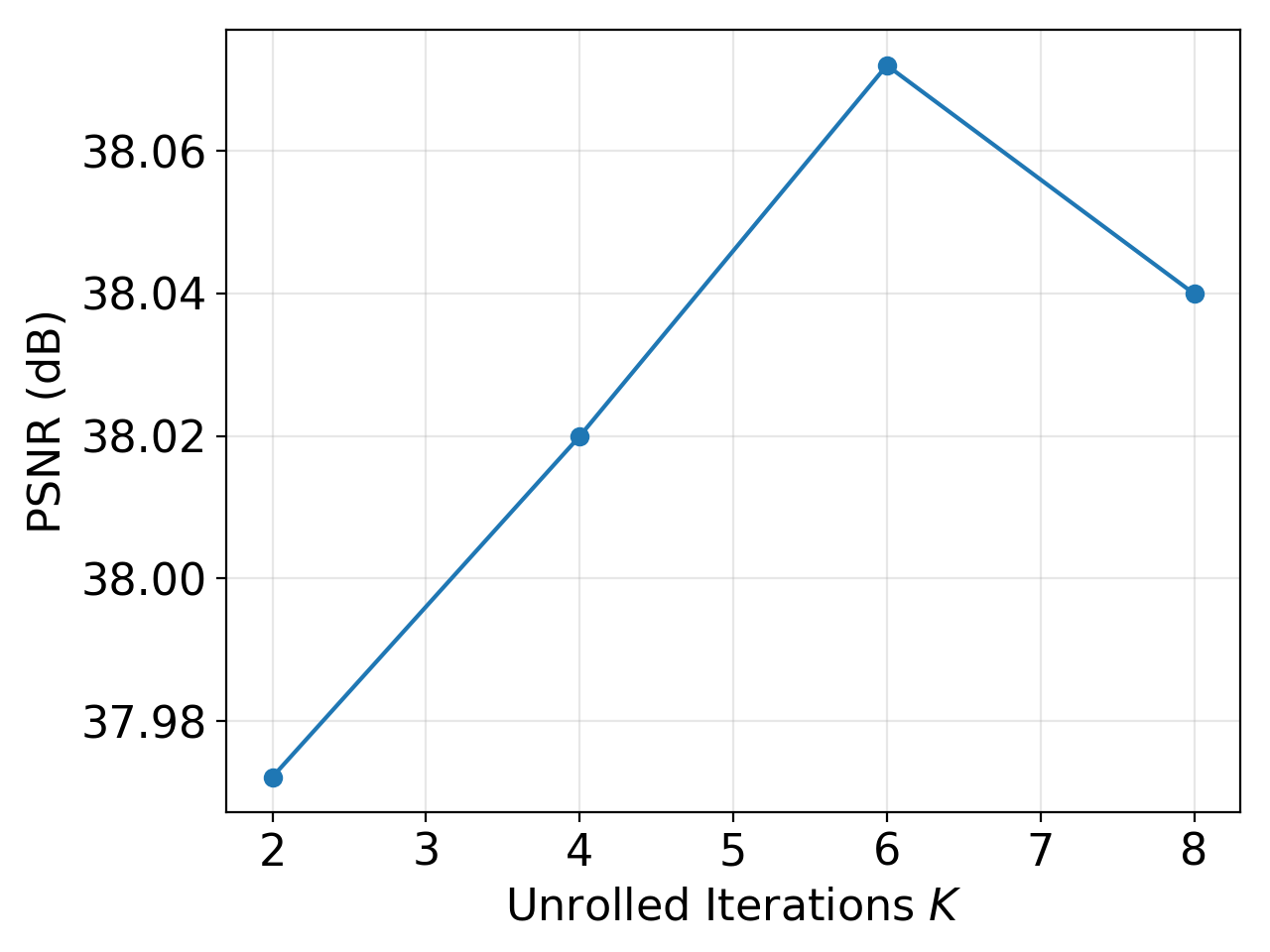}
    \end{subfigure}
    \hfill
    \begin{subfigure}[b]{0.48\linewidth}
        \centering
        \includegraphics[width=\linewidth]{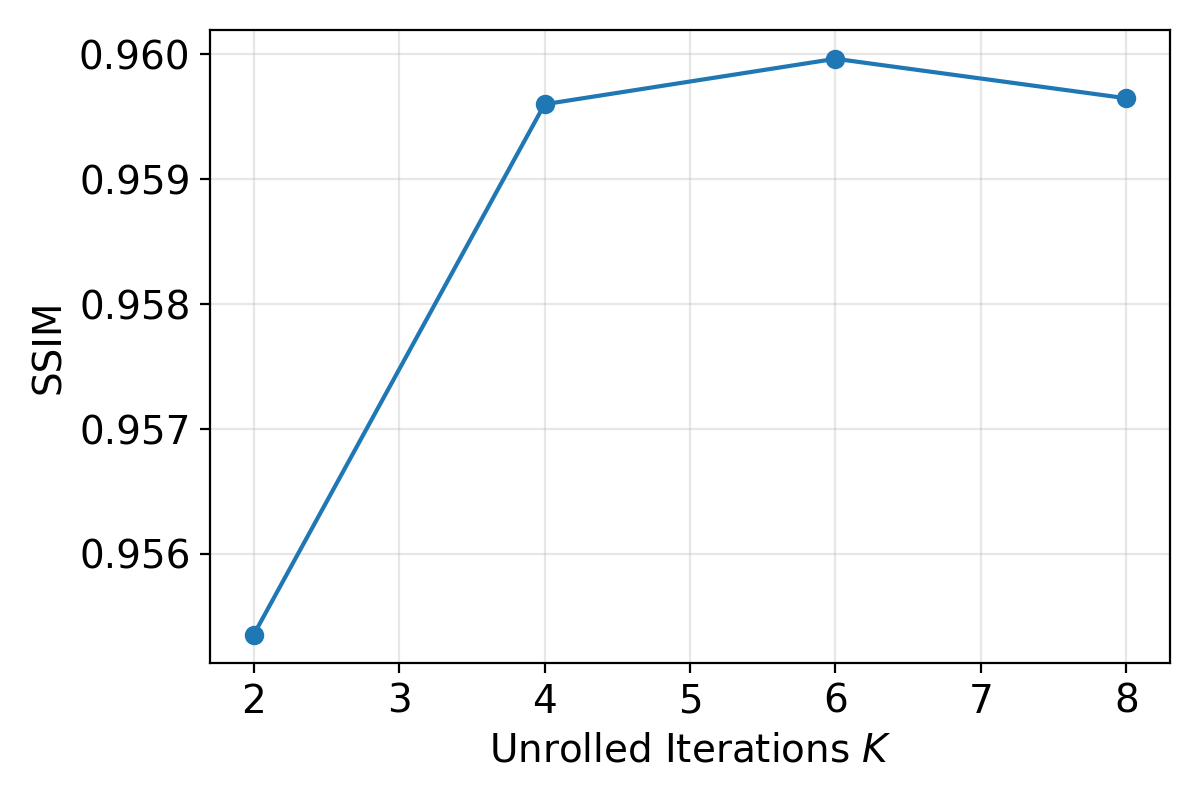}
    \end{subfigure}
    \caption{\, Sensitivity of MostNet reconstruction quality to the number of unrolled iterations $K$ (fixed $\lambda = 10^{-2}$). We get the best reconstruction quality at $K=6$ with the highest PSNR and SSIM at that point.}
    \label{fig:ablation_niter_both}
\end{figure}

We further assessed how the number of unrolled iterations, $K$, affects reconstruction quality for the proposed MostNet architecture. Here, $K$ denotes the number of unrolled stages, where each stage applies one alternating denoising and data-consistency (CG) update. As observed in Fig.~\ref{fig:ablation_niter_both}, reconstruction quality in terms of PSNR and SSIM improves consistently as $K$ increases from $2$ to $6$ (keeping $\lambda=10^{-2}$ and $\gamma=0.5$ fixed). Beyond this point, performance begins to degrade, while computational cost continues to increase approximately linearly.
Thus, $K = 6$ gives the best reconstruction quality among all the different cases in terms of PSNR and SSIM metrics.

\end{document}